\documentclass[sigconf, nonacm, pdfa]{acmart}

\usepackage{balance}
\usepackage{pifont}
\usepackage[linesnumbered,ruled,vlined]{algorithm2e}
\SetAlgoNlRelativeSize{0}
\usepackage{xcolor} 
\SetCommentSty{mycommstyle} 
\usepackage{adjustbox}
\usepackage{multirow}
\usepackage{booktabs}
\usepackage{enumitem}
\usepackage{multicol}
\usepackage[most]{tcolorbox}
\usepackage{amsthm} 
\usepackage{algorithmic}
\DontPrintSemicolon

\newtheoremstyle{myexamplestyle}
  {\topsep}   
  {\topsep}   
  {}          
  {}          
  {\bfseries} 
  {.}         
  {.5em}      
  {}          
\theoremstyle{myexamplestyle}
\newtheorem{example}{Example} 
\usepackage{amsmath} 
\usepackage{balance} 
\usepackage{booktabs} 
\usepackage{caption}
\usepackage{comment}
\usepackage{enumitem}
\setlist[itemize]{leftmargin=2.2em}

\usepackage{fancyvrb} 
\usepackage[symbol,flushmargin]{footmisc}
\usepackage{graphicx} 
\usepackage{makecell} 
\usepackage{multirow} 
\usepackage{soul}
\usepackage{graphicx}
\usepackage{subfigure} 
\usepackage{xcolor}
\usepackage{xspace} 
\usepackage{bm}
\usepackage{utfsym}
\usepackage{cancel}
\usepackage[linesnumbered,ruled,vlined]{algorithm2e}
\usepackage{pifont}
\usepackage{tikz}
\usepackage{marginnote}
\usepackage{mparhack}

\theoremstyle{definition}

\theoremstyle{plain}

\newtheorem{theorem}{Theorem}
\theoremstyle{remark}

\renewcommand{\paragraph}[1]{\vspace{0.1em}
\noindent\textbf{#1.}}
\renewcommand{\subparagraph}[1]{\vspace{0.1em}
\noindent\textit{\underline{#1.}}}

\usepackage{multicol}

\usepackage{wrapfig}
\usepackage{listings}
\usepackage[most]{tcolorbox}

\newtcolorbox[auto counter]{mybox}[1][]{%
breakable,
enhanced,
sharp corners,
colback=white,
fonttitle=\bfseries,
enlarge bottom at break by=5mm,
enlarge top at break by=5mm,
overlay first={%
    \draw[black, line width=0.5mm](frame.south west)--(frame.south east);
    \node[anchor=north east] at (frame.south east) {continued on next page};
    },
overlay middle={%
    \draw[black, line width=0.5mm](frame.south west)--(frame.south east);
    \draw[black, line width=0.5mm](frame.north west)--(frame.north east);
    \node[anchor=north east] at (frame.south east) {continued on next page};
    \node[anchor=south west] at (frame.north west) {continued from next page};
    },
overlay last={%
    \draw[black, line width=0.5mm](frame.north west)--(frame.north east);
    \node[anchor=south west] at (frame.north west) {continued from next page};},
#1
}

  {\list{}{\leftmargin=0.15in\rightmargin=0.15in}\item[]}%
  {\endlist}

\newtcolorbox{myquote}[1][]{
    colback=black!10,
    colframe=black!10,
    notitle,
    sharp corners,
    enhanced,
    breakable,
    left=2pt,
    right=2pt,
    top=2pt,
    bottom=2pt,
    ignore nobreak,
}

\definecolor{R1color}{HTML}{e66101} 
\definecolor{R2color}{HTML}{7b3294} 
\definecolor{R4color}{HTML}{018571} 

\usepackage{xcolor} 
\definecolor{revcolor}{HTML}{E66101}

\newtcolorbox{myquoteMeta}[1][]{
    colback=black!3,
    colframe=black!3,
    notitle,
    sharp corners,
    borderline west={1.5pt}{0pt}{blue},
    enhanced,
    breakable,
    left=2pt,
    right=2pt,
    top=2pt,
    bottom=2pt,
    ignore nobreak,
}

\newtcolorbox{myquoteR1}[1][]{
    colback=black!3,
    colframe=black!3,
    notitle,
    sharp corners,
    borderline west={1.5pt}{0pt}{R1color},
    enhanced,
    breakable,
    left=2pt,
    right=2pt,
    top=2pt,
    bottom=2pt,
    ignore nobreak,
}

\newtcolorbox{myquoteR2}[1][]{
    colback=black!3,
    colframe=black!3,
    notitle,
    sharp corners,
    borderline west={1.5pt}{0pt}{R2color},
    enhanced,
    breakable,
    left=2pt,
    right=2pt,
    top=2pt,
    bottom=2pt,
    ignore nobreak,
    #1
}

\newtcolorbox{myquoteR4}[1][]{
    colback=black!3,
    colframe=black!3,
    notitle,
    sharp corners,
    borderline west={1.5pt}{0pt}{R4color},
    enhanced,
    breakable,
    left=2pt,
    right=2pt,
    top=2pt,
    bottom=2pt,
    ignore nobreak,
}

\newif\ifextended\extendedfalse

  {\endminipage\par\vspace{10pt plus 2pt minus 5pt}}

\newcommand\vldbdoi{10.14778/3828612.3828613}
\newcommand\vldbpages{2536 - 2548}
\newcommand\vldbvolume{19}
\newcommand\vldbissue{10}
\newcommand\vldbyear{2026}
\newcommand\vldbauthors{\authors}
\newcommand\vldbtitle{\shorttitle} 
\newcommand\vldbavailabilityurl{https://github.com/Sakuraaa0/TVA.git}
\newcommand\vldbpagestyle{empty} 

\newcommand\dbname{\ensuremath{\textsf{TVA}}\xspace}

\begin{document}
\settopmatter{authorsperrow=4}
\title{\dbname: A Version-aware Temporal Graph Storage System for Real-time Analytics}



\author{Wenhao Li}
\affiliation{%
  \institution{\mbox{Renmin University of China}}
}
\email{ruclwh@ruc.edu.cn}

\author{Zhanhao Zhao}
\affiliation{%
  \institution{\mbox{Renmin University of China}}
}
\email{zhanhaozhao@ruc.edu.cn}

\author{Jinhao Dong}
\affiliation{%
  \institution{\mbox{Renmin University of China}}
}
\email{imonedjh@gmail.com}

\author{Jiamin Hou}
\affiliation{%
  \institution{\mbox{Zhejiang Univ. \& RUC}}
}
\email{jiaminhou@zju.edu.cn}

\author{Wei Lu}
\affiliation{%
  \institution{\mbox{Renmin University of China}}
}
\email{lu-wei@ruc.edu.cn}

\author{Yunhai Wang}
\affiliation{%
  \institution{\mbox{Renmin University of China}}
}
\email{cloudseawang@gmail.com}

\author{Xiaoyong Du}
\affiliation{%
  \institution{\mbox{Renmin University of China}}
}
\email{duyong@ruc.edu.cn}






\begin{abstract}
Analyzing temporal graphs can reveal valuable insights that are typically hidden in static graphs.
Unfortunately, existing graph storage systems either lack native temporal support or suffer from high latency when querying temporal graphs.
This paper presents \dbname, a new temporal graph storage system designed for efficient temporal query processing.
First, \dbname introduces a specialized multi-version storage architecture that separates version metadata from actual data, i.e., the property values associated with different versions of vertices and edges.
This architecture enables efficient version retrieval for a vertex or edge by quickly locating valid version metadata and directly dereferencing it to access the corresponding property values.
Second, we design tailored data structures, namely the temporal table and enhanced hopscotch-based hashing, to compactly organize the version metadata of adjacent vertices and edges, thus reducing random I/O for metadata lookups during the neighborhood scan initiated from a vertex.
Finally, to further accelerate neighborhood scans over multiple vertices, we propose a version-skipping strategy that reuses temporal information obtained from prior scans, thereby avoiding redundant metadata lookups across scans.
Empirical evaluations demonstrate that \dbname achieves up to $9.9\times$ lower temporal query latency and $2.2\times$ lower storage overhead compared to state-of-the-art temporal graph storage systems.



\end{abstract}

\maketitle

\pagestyle{\vldbpagestyle}
\begingroup\small\noindent\raggedright\textbf{PVLDB Reference Format:}\\
\vldbauthors. \vldbtitle. PVLDB, \vldbvolume(\vldbissue): \vldbpages, \vldbyear.\\
\href{https://doi.org/\vldbdoi}{doi:\vldbdoi}
\endgroup
\begingroup
\renewcommand\thefootnote{}\footnote{%
Wei Lu is the corresponding author.\\
This work is licensed under the Creative Commons BY-NC-ND 4.0 International License. Visit \url{https://creativecommons.org/licenses/by-nc-nd/4.0/} to view a copy of this license. For any use beyond those covered by this license, obtain permission by emailing \href{mailto:info@vldb.org}{info@vldb.org}. Copyright is held by the owner/author(s). Publication rights licensed to the VLDB Endowment. \\
\raggedright Proceedings of the VLDB Endowment, Vol. \vldbvolume, No. \vldbissue\ %
ISSN 2150-8097. \\
\href{https://doi.org/\vldbdoi}{doi:\vldbdoi} \\
}\addtocounter{footnote}{-1}\endgroup

\ifdefempty{\vldbavailabilityurl}{}{
\vspace{.3cm}
\begingroup\small\noindent\raggedright\textbf{PVLDB Artifact Availability:}\\
The source code, data, and/or other artifacts have been made available at \url{\vldbavailabilityurl}.
\endgroup
}

\section{Introduction}\label{sec:Introduction}



A \textit{temporal graph}, also known as a \textit{dynamic or time-varying graph}, represents the evolution of relationships between entities, the entities themselves, or both over time~\cite{DBLP:journals/pacmmod/XiaFL25,DBLP:conf/eurosys/HanMLWYZPCC14}. 
Unlike static graphs, temporal graphs retain rich temporal information, which is crucial for a wide range of applications, such as fraud detection~\cite{DBLP:journals/tkde/ChengWZZ22, DBLP:journals/pvldb/CaoYCZLQ19,DBLP:conf/sigmod/YeLHLS21} and social network analysis~\cite{DBLP:conf/eurosys/WilsonBSPZ09,DBLP:conf/sigmod/LometBMSWZ05,DBLP:journals/tos/MiaoHLWYZPCC15}.

\begin{figure}
  \centering
   \includegraphics[width=\linewidth]{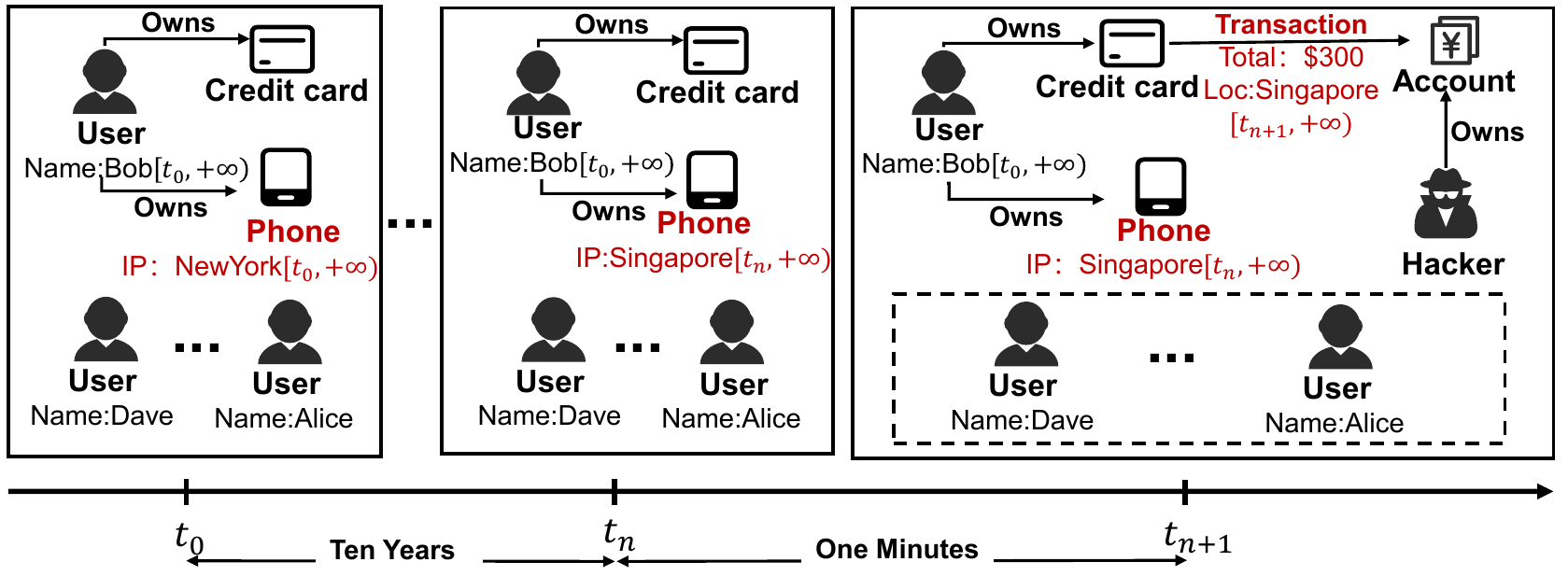}
  \caption{An Example of Financial Temporal Graph.}
  \label{fig:sample-intro}
\end{figure}

\begin{example}
\label{exp:intro}
Figure \ref{fig:sample-intro} shows a typical temporal graph in the financial domain, where vertices represent entities, such as users, credit cards, and bank accounts, and edges denote their relationships, such as ownership and transaction events. 
Each vertex and edge can have multiple versions over time.
Suppose that user Bob's phone IP address, stable in New York throughout the time interval [$t_0$, $t_{n}$), suddenly changes to Singapore at time $t_n$.
This IP change creates a new version of the vertex representing Bob’s phone.
Shortly afterward, at time $t_{n+1}$, Bob initiates a transaction in Singapore.
The pattern of a risky transaction following a sudden location change strongly suggests that Bob’s account has been compromised.
To detect such risks, one can leverage a fraud detection model with relevant graph data~\cite{DBLP:journals/tkde/XiangZCZ25,DBLP:journals/pvldb/XiaoW00O23,jiang2025dupin}.
However, if only a static graph (e.g., at time $t_{n+1}$) is provided, the model may fail to identify the fraud due to the absence of recent geolocation change information.
Therefore, it is essential to obtain a temporal graph within an appropriate time period, e.g., [$t_{n-1}$, $t_{n+1}$], that includes the necessary temporal context for effective fraud detection.
\qed
\end{example}

Continuing from Example~\ref{exp:intro}, a typical temporal query retrieves transactions that occur between [$t_{n-1}$, $t_{n+1}$], along with their corresponding user versions.
However, some graph systems~\cite{DBLP:conf/hpec/EdigerMRB12,DBLP:conf/fast/KumarH19,DBLP:journals/pvldb/FuchsGM22} retain only the current state of the graph, making them unable to handle such queries.
In contrast, systems like T-GQL~\cite{DBLP:journals/vldb/DebrouvierPPSV21} store versions as separate objects, incurring significant storage overhead.
Snapshot-based approaches, such as Clock-G~\cite{DBLP:conf/icde/MassriMPM22}, periodically store full snapshots of the graph and maintain logs of deltas between snapshots. It requires expensive reconstruction of snapshots from deltas.
A recent system, AeonG~\cite{DBLP:journals/pvldb/HouZWLJWD24}, applies fine-grained versioning but incurs redundant traversal and extensive random I/O when scanning version chains across large subgraphs. Furthermore, general-purpose storage systems are also ill-suited for temporal graphs: RDBMSs suffer from computationally expensive join operations during multi-hop traversals, while Key-Value (KV) stores lack structural awareness, leading to inefficient temporal range scans.

Addressing these challenges requires a new storage architecture that can efficiently locate relevant versions of a given time condition.
Our goal is to minimize the version traversal cost during temporal query processing, thereby improving the overall performance.
However, achieving this requires addressing three key challenges:
\ding{182} First, each vertex or edge may contain multiple properties, and updates typically affect only a subset of them.
To reduce storage overhead, systems often adopt delta-based encoding~\cite{DBLP:journals/pvldb/HouZWLJWD24,DBLP:conf/icde/MassriMPM22}, where only the modified properties are stored in each version.
However, this makes it infeasible to reconstruct a complete version without traversing its version chain.
As a result, achieving fast version retrieval with minimal storage redundancy is not straightforward.
\ding{183} {Second, a vertex can be connected to multiple edges, each maintaining its own version chain. Existing data structures are primarily designed and optimized for accessing the latest version of data, while historical versions are often appended to each data item using a simple linked structure~\cite{DBLP:journals/pvldb/HouZWLJWD24,DBLP:conf/icde/MassriMPM22,DBLP:journals/vldb/DebrouvierPPSV21}. This design prevents historical data queries from benefiting from the core optimization mechanisms of the system. Designing a unified structure to co-manage these version chains efficiently, particularly for high-degree vertices, remains a significant challenge.
\ding{184} Third, temporal queries typically involve scanning the neighborhoods of multiple vertices~\cite{DBLP:conf/sigmod/MalewiczABDHLC10,DBLP:journals/tos/MiaoHLWYZPCC15}.
However, existing systems process each of these neighborhood scans independently. Leveraging this cross-query temporal locality to avoid redundant version traversals and computation is key to enhancing overall query efficiency.
In this paper, we present \dbname, a version-aware graph storage system designed for efficient temporal query processing. 
We introduce a multi-version storage architecture that separates version metadata from actual data. 
The actual property values of graph objects are stored in a compact structure, with pointers in the metadata, allowing retrieval by first locating the metadata and then dereferencing it. 
This minimizes random I/O and enables low-latency version retrieval (Challenge~\ding{182}).
We also design specialized data structures: a temporal table for compact vertex metadata and a hopscotch-based hashing structure for topological relationships. 
We use a hopscotch-inspired algorithm to co-locate temporally related metadata in the same hash bucket, storing versions in temporal order with a bounded offset. 
This allows logarithmic lookup time for version metadata, while maintaining access locality by dynamically relocating hot and cold data based on update frequency. SIMD techniques further accelerate version evaluation (Challenge~\ding{183}).
Finally, we propose a version-skipping approach that links versions within a snapshot, enabling subsequent scans to resume from previously retrieved metadata, avoiding redundant lookups (Challenge~\ding{184}).

In summary, we make the following contributions:

\vspace{-4mm}
\begin{itemize}[leftmargin=0.3cm]

\item We propose \dbname, a new temporal graph storage system designed for efficient temporal query processing. 

\item We design a multi-version storage architecture tailored for temporal graphs, which reduces redundant version traversal cost without sacrificing storage efficiency.

\item We introduce a hopscotch-based hash structure to enable bounded lookup time when performing neighborhood scans under given time conditions.
In addition, we utilize SIMD technology to enable parallel query processing.

\item We propose a version-skipping approach that reuses temporal information from previous scans to improve temporal query performance.


\item We conduct extensive experimental evaluations on widely-used benchmarks, demonstrating that \dbname achieves up to $2.2\times$ lower storage overhead and up to $9.9\times$ faster temporal query processing compared to state-of-the-art graph systems.
\end{itemize}
\section{Background}
\label{sec:PRELIMINARIES}
In this section, we review existing temporal graph storage systems, discuss the basic techniques used in the design of \dbname, and formally define the problem that \dbname aims to address.

\subsection{Temporal Graph Model}
\label{sec:temporal_graph_model}

\textbf{Temporal Graph Definition.} A temporal graph is defined as $\mathcal{G} = (\mathcal{X}, \mathcal{E})$:
\begin{itemize}[leftmargin=0.3cm]
\item Each vertex $x \in \mathcal{X}$ is associated with a set of properties $\rho$ and an active time interval $\tau = [t_{\text{start}}, t_{\text{end}})$.
\item Each edge $e \in \mathcal{E}$ is a tuple $(src, dst, \tau, \rho)$, where $src, dst \in \mathcal{X}$ are the endpoints of the edge, $\tau = [t_{\text{start}}, t_{\text{end}})$ denotes the edge's active time interval, and $\rho$ is a set of properties.
\end{itemize}
Each vertex and edge may possess multiple labels. To simplify the subsequent discussion, we assume that each vertex or edge has a single label. For example, we denote a user vertex as $x^{\text{user}}$, a phone vertex as $x^{\text{phone}}$, and a transaction edge as $e^{\text{txn}}$. A one-to-one pair (e.g., user-phone) is denoted as $(x^{\text{user}}, x^{\text{phone}})$.
We denote the different versions of a vertex (edge) as $x_n.v_m$ ($e_n.v_m$), where $x_n$ (or $e_n$) refers to the $n$-th vertex (or edge), and $v_m$ indicates its $m$-th version.
Throughout this paper, both vertices and edges are collectively referred to as graph objects. For versioning semantics, TVA supports both logical version numbers (e.g., transaction IDs) and physical timestamps. Without loss of generality, we use physical time intervals $[t_{start}, t_{end})$ in our examples.


\noindent
\textbf{Temporal Graph Operation.}
Temporal graph updates evolve the state of the graph while preserving historical information.
The supported operations are as follows:

\begin{itemize}[leftmargin=0.3cm]
    \item \textbf{Create Vertex or Edge:} Add a graph object \(o\) to the temporal graph $\mathcal{G}$ and assign corresponding label and properties, with a time version assigned as $\tau = [t_1, +\infty)$.
    \item \textbf{Delete Vertex or Edge:} Find the graph object \(o \in \mathcal{X} \cup \mathcal{E}\) of the corresponding version, update its $\tau = [t_1,+\infty)$ to $[t_1, t_2)$ to mark its deletion.
    \item \textbf{Update Vertex or Edge:} Find the graph object \(o \in \mathcal{X} \cup \mathcal{E}\) of the corresponding version, update its $\tau = [t_1, +\infty)$ to $[t_1, t_2)$ and mark it as a historical version. Then, create a new version with $\tau = [t_2, +\infty)$ to represent the latest version and apply the necessary updates to the vertex or edge.
\end{itemize}


\noindent\textbf{Temporal Graph Query.}
Temporal graph queries aim to retrieve graph objects that satisfy both a temporal condition and a predicate over their attributes.
Given a time point or interval $t_q$, a temporal query returns all vertices and edges whose validity intervals intersect $t_q$. The result can be viewed as a static graph snapshot obtained by filtering out objects that are not valid during $t_q$.
Formally, let a graph object $o \in \mathcal{X} \cup \mathcal{E}$ have a unique identifier $o.id$ and a validity interval $o.\tau$. A temporal query is defined as:
\begin{equation}\label{func:find_all}
            O_{\text{query}} = \left\{o \in \mathcal{X} \cup \mathcal{E} \;\middle|\; P(o) = \text{true} \land\; o.\tau \cap t_q \neq \varnothing \right\},
\end{equation}
where $P(o)$ is a predicate over graph objects.
This formulation generalizes common temporal query types. For example, retrieving a specific object corresponds to $P(o): o.id = id_q$; and retrieving a set of objects corresponds to a general predicate, e.g., $o.\rho[\text{Location}] = \text{``New York''}$.

\subsection{Hopscotch Hash}\label{sec:2.2Hop}

\begin{figure}
  \centering
  \includegraphics[width=\linewidth]{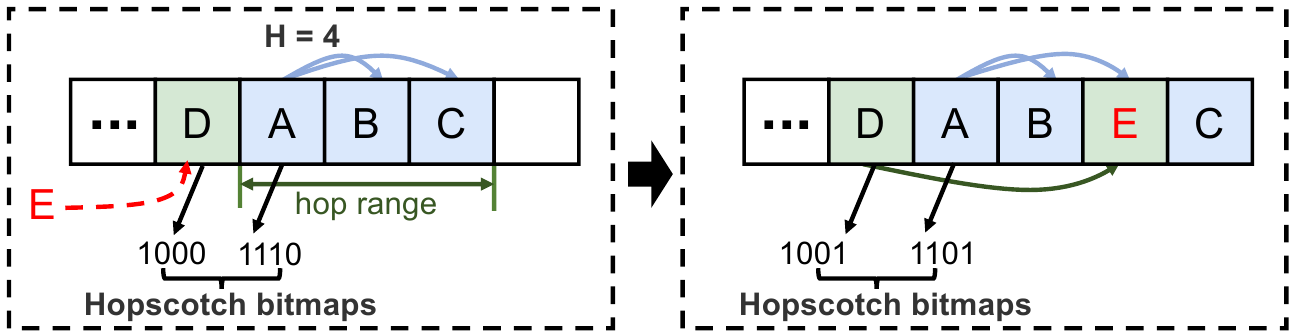}
  \caption{An Example of Hopscotch Hash Table. }
  \Description{A Hopscotch hash table with a Maximum jump distance of 3.}
  \label{fig:sample-hopscotch_hash_table}
\end{figure}


Hopscotch Hash~\cite{DBLP:conf/wdag/HerlihyST08} is an open-addressing hash algorithm that constrains each key to a fixed-size neighborhood (\(\mathcal{H}\)) near its primary slot. 
Each slot maintains an \(\mathcal{H}\)-bit bitmap, indicating which of the next \(\mathcal{H}\) slots hold its key. 
During insertion, if a key's primary slot is occupied, the algorithm identifies the first available empty slot. If this empty slot is outside the desired neighborhood, the algorithm searches the previous \(\mathcal{H}\)-1 items to find an item that can be swapped into the empty slot without violating its neighborhood constraint. The swap is repeated until an empty slot is moved into the neighborhood of the key's primary slot.
Figure~\ref{fig:sample-hopscotch_hash_table} shows an example of this process.  
In the left subfigure, keys A, B, and C share the same hash value and are inserted into slots within the hop range of A's primary slot (Bitmap=1110). Similarly, D is placed in its own primary slot (Bitmap=1000).  
When inserting E, which has the same hash value as D, its primary slot is full. Through the hopscotch insertion process, E is moved into a slot (Bitmap=1001) that still falls within the hop range of D’s primary slot.

Although the original hopscotch insertion algorithm is effective in resolving hash collisions, it fails to preserve data ordering and is not designed for managing temporal data versions. Instead of a direct application, we innovatively extend and redesign the hopping mechanism to propose an enhanced hopscotch-hash algorithm, which is specially tailored for organizing multiple temporal versions of data objects within a unified hash table. Our design addresses the specific challenges of temporal locality and version lookup efficiency, as detailed in §~\ref{Hopscotch Hash}.

\begin{figure*}
  \centering
  \includegraphics[width=0.95\linewidth]{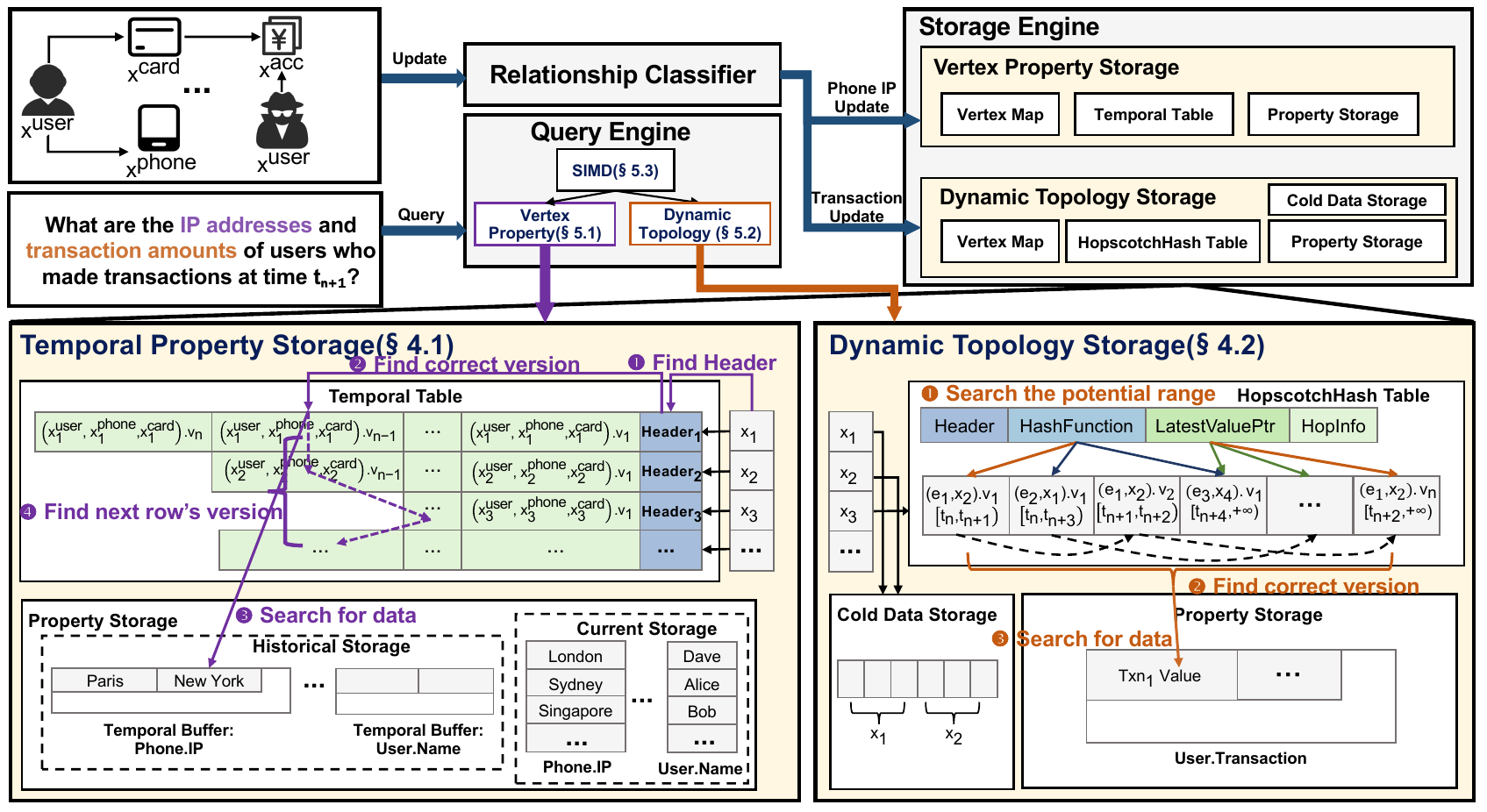}
  \caption{Overall Architecture of \dbname.}
  \Description{Overall architecture of \dbname.}
  \label{fig:overview.pdf}
\end{figure*}

\section{System Overview}\label{sec:BACKGROUND AND MOTIVATIONS}

\dbname supports efficient temporal graph management through two key components:
the storage engine and the query engine.

\subsection{Storage Engine}
As shown in Figure~\ref{fig:overview.pdf}, \dbname's storage engine adopts a hybrid architecture with two specialized components, designed to handle the different types of data in a temporal graph. 
An incoming update is routed by the \texttt{Relationship Classifier} to the Temporal Property Storage if it modifies vertices' intrinsic properties, or to the Dynamic Topology Storage if it creates, deletes, or modifies an edge between vertices. 
The central idea behind these two components is to separately manage version metadata and the actual data.


\noindent\textbf{Temporal Property Storage:}
This component is optimized for efficient management of vertex properties (e.g., a user's name or a phone's IP address). At any moment, each vertex maintains at most one current value per property; therefore, a columnar storage layout is employed to facilitate efficient access to current data~\cite{DBLP:conf/sigmod/AbadiMH08}. Historical values for each property are stored separately in a specialized structure called \texttt{Temporal Buffer}. Versioning metadata is maintained by \texttt{Temporal Table} structure. 
Upon property updates, the existing value is not overwritten; instead, it is transferred to the corresponding \texttt{Temporal Buffer}, and a new record is added to the \texttt{Temporal Table} to index this update. 
We will provide further details in §~\ref{column with td}.

\noindent\textbf{Dynamic Topology Storage:} This component is responsible for managing the dynamic connections between vertices (e.g., transactions or social links), where a vertex can have multiple edges for the same relationship. 
Such dynamism makes columnar layouts for current data, where data location can be calculated by a predictable offset~\cite{raasveldt2019duckdb}, unsuitable. 
We observe that for the edges of the same vertex, there is no strong order correlation among different edges, but for different versions of the same edge, it is best to store them in chronological order to speed up lookup. 
To achieve this, we construct a storage structure based on a hopscotch algorithm.
This structure stores all edges of a vertex within a single hash bucket, and it organizes all versions of an edge chronologically within a ``neighborhood''. In addition, we optimize dynamic topology storage through a cold-hot data separation strategy. 
We will introduce the details in §~\ref{Hopscotch Hash}.

\subsection{Query Engine}

The query engine processes user queries and retrieves relevant graph data from the hybrid storage. \dbname supports all temporal graph operations described in §~\ref{sec:temporal_graph_model}.

For Temporal Property Storage, all basic metadata for each vertex is stored in the \texttt{Header} of the \texttt{Temporal Table}. \dbname first identifies the target vertex, then locates its \texttt{Header} in the \texttt{Temporal Table} based on the offset to check whether the data exists in the \texttt{Current Storage}. If present, the data can be read directly from the corresponding column; if not, \dbname performs a binary search to locate the correct \texttt{Version} in the \texttt{Temporal Table}, using the offset recorded in that \texttt{Version} to determine the data position. Additionally, when a query retrieves all data from the same timestamp, \texttt{NextOffset} structure is introduced to use information from the previous row, thereby accelerating queries for subsequent rows. Further details are provided in §~\ref{sec:1-1_QUERY_ENGINE}.


For Dynamic Topology Storage, we similarly start by identifying the relevant vertex. Subsequently, using the \texttt{HashFunction} and \texttt{LatestValuePtr} in the storage structure pointed to by that vertex, we determine the potential range for the edge we need to find. 
Finally, we locate the specific data within this range. 
Further details are provided in §~\ref{sec:n-n_QUERY_ENGINE}.

These two structures enable efficient temporal queries on both properties and topology. 
In addition, these compact storage layouts enable us to efficiently leverage SIMD hardware acceleration, thereby further speeding up temporal queries. 
More details are provided in §~\ref{sec:simd}.

\begin{example}
Figure~\ref{fig:overview.pdf} presents example query statements that demonstrate operations on both storages. The query highlighted in purple corresponds to Temporal Property Storage, while the one in orange illustrates an operation on Dynamic Topology Storage.

The purple query corresponds to: ``Snapshot of IP addresses at $t_{n+1}$.'' It locates the first vertex $x_1$ and performs a binary search on the Temporal Table to find the matching version. For subsequent queries, the previously identified \texttt{Version} can be leveraged to quickly locate potential neighboring position of the next row's \texttt{Version}, thereby narrowing the search. The orange query resolves the question, ``What are the transaction amounts at time $t_{n+1}$?'' We take Bob’s transaction as an example. Let Bob’s vertex be $x_3$, and the relevant edge be $e_1$. Using the \texttt{HashFunction}, we can locate the oldest version of $e_1$, while \texttt{LatestValuePtr} points to its latest version. Searching within this range then retrieves the desired data. If we need to search for the next edge of \( x_3 \), we can similarly utilize the information obtained in $e_1$.
\qed
\end{example}

\section{Storage Design}\label{sec:SYSTEM DESIGN}

In this section, we detail the design of the Temporal Property Storage and dynamic topology storage. We provide detailed pseudocode for Create, Read, Update, and Delete (CRUD) operations to explicitly demonstrate how TVA handles data manipulation.

\subsection{Temporal Property Storage}\label{column with td}

\textbf{Key Idea.}
We propose a multi-version storage architecture for vertex properties, based on the observation that each vertex maintains at most one current value for any given property. The current state is stored in a columnar layout, which allows efficient direct access for current queries.
To address the inefficiency of traditional methods that require traversing version chains for historical queries, we introduce a dedicated historical storage design. The core of this design is the \texttt{Temporal Table}, which separates version metadata from property values and stores all historical data in a unified, compact region. Metadata entries are stored in a tightly packed, contiguous format to enable rapid direct lookups. As a result, retrieving any version requires only locating its metadata and dereferencing the corresponding value, thereby minimizing overhead.
In addition, we introduce \texttt{NextOffset} structure within the \texttt{Temporal Table} that links temporally-related versions across different vertices, accelerating snapshot queries that require data from the same timestamp. 

\begin{figure}[b]
\centering
  \includegraphics[width=\linewidth]{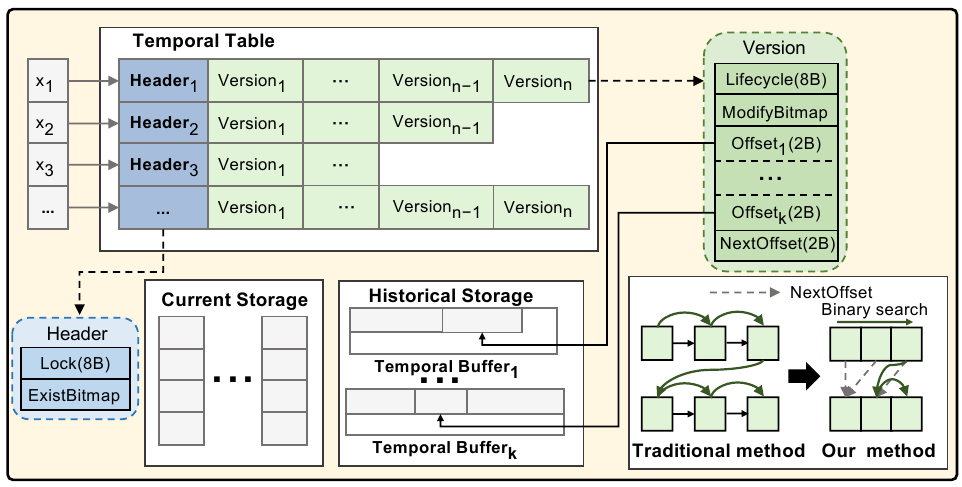}
  \caption{The Storage Format for Vertex Properties.}
  \Description{The Storage Format for Vertex Properties.}
  \label{fig:column.pdf}
\end{figure}

\noindent\textbf{Storage Format.} As illustrated in Figure \ref{fig:column.pdf}, the fundamental layout of \texttt{Current Storage} in this component is columnar storage. The vertex IDs allocated according to the chosen scheme serve as primary keys. Properties are organized into distinct columns and indexed by their labels and property keys. 
To manage temporal data and updates effectively, we introduce two critical extensions:

\texttt{Temporal Table}: For each vertex ID, \texttt{Temporal Table} manages version metadata and stores it separately from the corresponding property values. Each row consists of a \texttt{Header} and several \texttt{Version} entries. The \texttt{Header} contains a \texttt{LOCK} field for concurrency control during updates, as well as an \texttt{ExistBitmap}, which indicates which property columns currently have data for the corresponding vertex. Each \texttt{Version} entry represents an update to the vertex and records the historical versions of the relevant properties. Specifically, it includes a \texttt{Lifecycle} field that specifies the validity interval $[t_{\text{start}}, t_{\text{end}})$ of that historical state, and a \texttt{ModifyBitmap} that identifies which property columns are involved in this version. Additionally, an array of \texttt{Offset\(_1\)} to \texttt{Offset\(_k\)} is included, where $k$ denotes the number of property columns. Each \texttt{Offset\(_j\)} points to the storage location of the historical value for the $j$-th property within the \texttt{Temporal Buffer}; if it is set to \texttt{nullptr}, it instead refers to the value in the \texttt{Current Storage}. Finally, the \texttt{Version} structure contains a \texttt{NextOffset} pointer, which links to a specific \texttt{Version} in the next row. When a new \texttt{Version} is created, this pointer is set to the latest \texttt{Version} in the following row at that time. These \texttt{NextOffset} pointers collectively form multiple version chains, through which we can efficiently retrieve subsequent data points within the same temporal context.

 \texttt{Historical Storage:} This is a large, append-only region where the actual historical values are stored. Each property has its corresponding \texttt{Temporal buffer}. When a property is updated, its previous value is written contiguously into its buffer.

\begin{algorithm}[t]
\small

\caption{Create and Update Operations in Temporal Property Storage}\label{alg:crud_operations}

\KwIn{Vertex $vid$, PropKey $p$, Value $v$, Time $t$, Temporal Table $\mathcal{T}$, Current Store $\mathcal{C}$, Historical Store $\mathcal{H}$}
\LinesNumbered
\SetKwFunction{FCreate}{Create}
\SetKwFunction{FUpdate}{Update}
\SetKwFunction{FNewVersion}{NewVersion}
\SetKwFunction{FAppend}{Append}
\SetKwProg{Fn}{Function}{:}{\KwRet}

\Fn{\FCreate{$vid,\; p,\; v,\; t$}}{
    $\mathcal{C}[vid][p] \leftarrow v$\\
    $ver \leftarrow$ \FNewVersion{$t,\; \infty$}\\
    $ver.Offset[p] \leftarrow \textbf{null}$\\
    $\mathcal{T}[vid].$\FAppend{$ver$}\\
}
\Fn{\FUpdate{$vid,\; p,\; v_{\text{new}},\; t$}}{
    $v_{\text{old}} \leftarrow \mathcal{C}[vid][p]$\\
    $\mathcal{C}[vid][p] \leftarrow v_{\text{new}}$\\
    $offset_{\text{old}} \leftarrow \mathcal{H}.$\FAppend{$v_{\text{old}}$}\\
    $ver_{\text{archived}} \leftarrow$ \FNewVersion{$t$}\\
    $ver_{\text{archived}}.Offset[p] \leftarrow offset_{\text{old}}$\\
    $\mathcal{T}[vid].$\FAppend{$ver_{\text{archived}}$}\\
}
\end{algorithm}

\noindent\textbf{Insert, Update or Delete.}
We design the system to maintain current data efficiently while archiving historical versions without costly reorganizations. The core idea is to physically separate current and historical states. In our system, a delete operation is treated as a special type of update operation and how read operations are performed will be elaborated in Section~\ref{sec:TEMPORAL_QUERY_ENGINE}. To explicitly demonstrate how \dbname handles data manipulation, we present the pseudocode for Graph operations in Algorithm \ref{alg:crud_operations}.
\\
For \textbf{Create} operations (lines 1-5), \dbname directly writes the value to the \texttt{Current Storage} to ensure $O(1)$ access for latest-state queries, while initializing the version metadata in the \texttt{Temporal Table}.
\\
The \textbf{Update} process (lines 6-12) is the core of our version-aware design. When vertex properties are updated at time $t$, we proceed as follows:
First, to prioritize the performance of reading the latest graph state, the new property value ($v_{\text{new}}$) is written directly into the corresponding property column of the \texttt{Current Storage} (line 8), implicitly marking its validity as $[t, +\infty)$.
Simultaneously, the previous property value ($v_{\text{old}}$) is preserved: it is appended to the append-only \texttt{Historical Storage} (\textit{Temporal Buffer}), as shown in line 9.
Finally, a new \texttt{Version} entry is created and appended to the \texttt{Temporal Table} to manage the metadata of this state change (lines 10-12). This new entry records the lifecycle end time ($t$) of the old version (representing the interval $[t_{\text{current}}, t)$) and stores the offset pointing to the archived $v_{\text{old}}$. Additionally, as detailed in Section 4.1, this new version entry maintains a \texttt{NextOffset} pointer, linking it to the latest version of the subsequent vertex to accelerate cross-vertex snapshot scanning.

\subsection{Dynamic Topology Storage}\label{Hopscotch Hash}


\textbf{Key Idea.}
For dynamic topology and its associated edge properties, it is difficult to use a simple vertex-to-storage mapping because one vertex can be linked to many data items of the same relationship. Additionally, the uncertainty in the number of edges associated with each vertex makes it difficult to predict the storage requirements needed for compact storage. This issue is even more obvious in real graphs, where some vertices have many edges and are updated often~\cite{DBLP:conf/sigmod/ZhangCWWYZZCZHC24,DBLP:journals/pacmmod/YuGTSZYLZLLYZ24}, leading to a large number of old versions.
To solve this, we use a mixed storage method: infrequently updated (``cold'') data is stored in a \texttt{SparseArray}, while frequently updated (``hot'') data is managed by an improved \texttt{HopscotchHash Table} that is designed for handling time-related data. This structure stores all objects with the same relationship in the same hash bucket. Our proposed ``Hopscotch'' algorithm guarantees that all historical versions of a given data item are stored in temporal order with a bounded offset.

\begin{figure}
\centering
  \includegraphics[width=\linewidth]{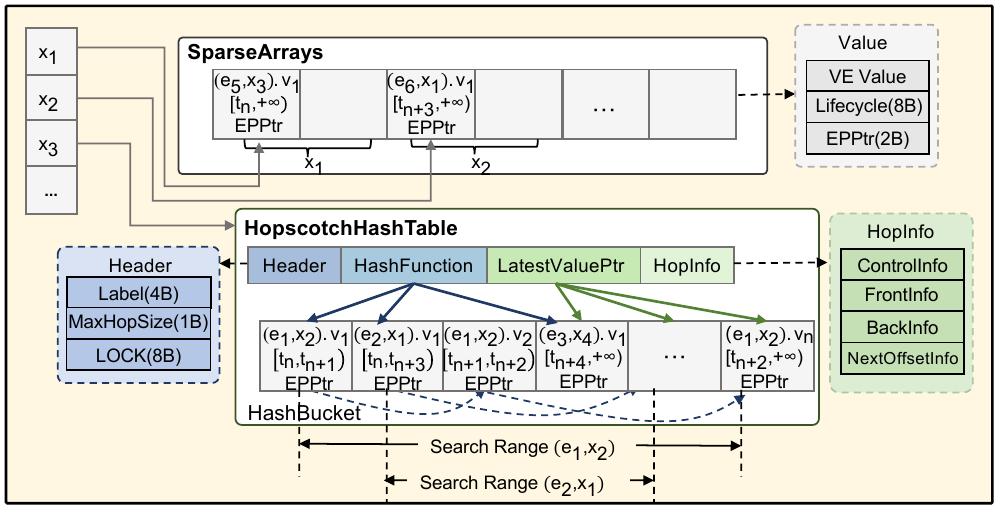}
  \caption{The Storage Format for Dynamic Topologies.
  }
  \Description{The Storage Format for dynamic topology data, which is organized by a HashTable with Hopscotch algorithm.}
  \label{fig:detaiHopHash.pdf}
\end{figure}

\noindent\textbf{Storage Format.}
The overall architecture is shown in Figure \ref{fig:detaiHopHash.pdf}.
The system stores data based on vertex degrees using two structures: \texttt{SparseArray} for ``cold'' data storage and \texttt{HopscotchHash Table} for ``hot'' data storage.
We define concrete criteria for this division: A vertex $v$ is classified as ``hot'' and migrated to the HopscotchHash Table if its degree exceeds a threshold $T_{deg}$, or its total version count reaches $T_{ver}$, formally expressed as $\sum_{e \in E(v)} N_v(e) \ge T_{ver}$ (where $N_v(e)$ is the version count of edge $e$). This migration is triggered asynchronously to minimize blocking. A background worker allocates a new hash bucket, re-organizes the existing data, and atomically updates the pointer, allowing the current write operation to complete without blocking. Conversely, migration from Hopscotch back to SparseArray is rare; it only occurs during a vacuum/purge operation where data prior to a specific timestamp is cleared, reducing the edge count below the threshold.
A \texttt{Vertex Map} directs different vertices to their respective storage units. Both structures employ a unified minimal storage unit for edge information that encapsulates three components: \texttt{VE Value} $(e_i, x_j)$, which captures the graph's topological information; \texttt{Lifecycle}, specifying the validity interval $[t_{\text{start}}, t_{\text{end}})$ of that historical state; and \texttt{EPPtr}, which points to the specific storage location of the edge's corresponding property. Next, we will introduce these two specific data structures.

 \texttt{SparseArray:} For ``cold'' data storage, it allocates fixed-size data slots for each vertex. This array-based structure offers good spatial locality, facilitating efficient scanning operations. Edges are inserted in a temporally ordered manner within a vertex's allocated space, simplifying subsequent time-series queries for that vertex's edges. When the number of elements reaches its maximum limit, all data will be transferred to the \texttt{HopscotchHash Table} for storage.
 
 \texttt{HopscotchHash Table:} For ``hot'' data storage, which are associated with numerous and frequently updated edges, HopscotchHash Table is employed. This structure is designed for efficient insertion and \(O(1)\) retrieval of the latest edge data, with historical versions accessible with logarithmic time complexity. The architecture of the HopscotchHash Table integrates several components: \texttt{Header}, which maintains metadata such as the label, the \texttt{MaxHopDistance} parameter, and \texttt{Lock} for concurrency control; \texttt{HashFunction} that maps edges to 64-bit hash values, while additionally recording the upper seven bits in the \texttt{HopInfo} to enhance SIMD query operations, as detailed in §~\ref{sec:simd}. The position indicated by \texttt{HashFunction} can be considered as the initial location of the corresponding item. To track the most recent version of each item, a dedicated hash table, \texttt{LatestValuePtr}, indexes data item identifiers to the storage location of their latest versions in the \texttt{HashBucket}.

Version control and efficient traversal are further enabled by four arrays collectively referred to as \texttt{HopInfo}. \texttt{ControlInfo} stores the upper seven bits of the hash value for each \texttt{HashBucket} slot. 
\texttt{FrontInfo} and \texttt{BackInfo} record bidirectional version offsets, respectively tracking links to the chronologically preceding and succeeding versions of a given entry. \texttt{MaxHopDistance} guarantees that all versions of a particular edge are stored within a bounded offset. In addition, the \texttt{NextOffset} array facilitates fast queries for subsequent data entries sharing the same timestamp.

\noindent\textbf{Insert, Update, or Delete.} 
These graph operations require finding the corresponding data insertion position in the Hopscotch Hash Table. 
Therefore, the key challenge is to resolve conflicts between temporal records.
To address this issue, we propose a novel Hopscotch algorithm for temporal data storage.

\noindent\textbf{Two Hopscotch Ways.}
When inserting a new version of a data item, we first obtain the most recent position $p$ via \texttt{LatestValuePtr}. The new version must be inserted within the range $[p+1, p+\texttt{MaxHopDistance}]$. If no empty slot is found within this range, the system must create a free slot via a ``Hopscotch'' algorithm, which involves relocating existing data items.



\begin{algorithm}[t]
\small
\caption{Hop Update Process}\label{alg:update_hop}
\KwIn{Free index $idx_{\text{free}}$, Destination index $idx_{\text{dst}}$, FrontInfo $fronts$, BackInfo $backs$, Max hop distance $d_{\text{max\_hop}}$, Judge distance $d_{\text{judge}}$, HashBucket $\mathcal{B}$}
\KwOut{An updated HopscotchHash Table with available free slot}
\LinesNumbered
\SetKwFunction{FUpdateHop}{UpdateHop}
\SetKwFunction{FHopChain}{HopChain}
\SetKwFunction{FInHopRange}{InHopRange}
\SetKwFunction{FHopStep}{HopStep}
\SetKwFunction{FFindInitial}{FindInitial}
\SetKwFunction{FIsInitial}{IsInitial}
\SetKwFunction{FUpdateInfos}{UpdateInfos}
\SetKwFunction{FSwap}{Swap}
\SetKwProg{Fn}{Function}{:}{\KwRet}

\Fn{\FUpdateHop{$idx_{\text{free}},\; idx_{\text{dst}}$}}{
    $idx_{\text{now}} \leftarrow idx_{\text{free}} - 1$ \\
    \While{$idx_{\text{free}} \notin [idx_{\text{dst}},idx_{\text{dst}} +d_{\text{max\_hop}} ]$ }{
            \If{$idx_{\text{free}} - idx_{\text{dst}} \leq d_{\text{judge}}$}{ 
                \eIf{$d_{\text{max\_hop}}-fronts[idx_{\text{now}}] \ge idx_{\text{free}} - idx_{\text{now}} $}{
                     \FSwap{$idx_{\text{now}}$ , $idx_{\text{free}}$}\\
                     $idx_{\text{free}} \leftarrow idx_{\text{now}}$\\
                     $idx_{\text{now}} \leftarrow idx_{\text{free}} - 1$\\
                }{
                  $idx_{\text{now}} \leftarrow idx_{\text{now}} - 1$\\
                }
            }
            \Else{
                    $idx_{\text{initial}} \leftarrow idx_{\text{now}}$\\
                    \While{ $fronts[idx_{\text{initial}} ]\neq 0$}{
                         $idx_{\text{initial}} \leftarrow idx_{\text{initial}} - fronts[idx_{\text{initial}}]$\\
                    }
                    \eIf{$idx_{\text{initial}} \geq idx_{\text{dst}}$}{
                        $idx_{\text{free}} \leftarrow$ \FHopChain{$idx_{\text{now}}$, $idx_{\text{free}}$}\\
                        $idx_{\text{now}} \leftarrow idx_{\text{free}} - 1$\\
                    }{
                        $idx_{\text{now}} \leftarrow idx_{\text{now}} - 1$\\
                }
            }
    }
}

\Fn{\FSwap{$i,\; j$}}{
    $\mathcal{B}[j] \leftarrow \mathcal{B}[i]$ \\
    $backs[i - fronts[i]],\; fronts[j] \leftarrow fronts[i] + j - i$\\
    $fronts[i+backs[i]],\; backs[j] \leftarrow backs[i] - j + i$\\
}

\Fn{\FHopChain{$idx_{\text{end}},\; idx_{\text{free}}$}}{
    $idx_{\text{right}} \leftarrow idx_{\text{free}}$\\
    $idx_{\text{left}} \leftarrow idx_{\text{end}}$\\
    \While{ $fronts[idx_{\text{left}} ]\neq 0$}{
         \FSwap{$idx_{\text{left}}$ , $idx_{\text{right}}$}\\
        $idx_{\text{right}} \leftarrow idx_{\text{left}}$\\
        $idx_{\text{left}} \leftarrow idx_{\text{left}} - fronts[idx_{\text{left}}]$\\
    }
    \KwRet $idx_{\text{left}}$\\
}

\end{algorithm}

In practice, to efficiently perform the operations illustrated in the example, we employ two distinct strategies based on the distance between the target location and the nearest empty slot.


As shown in Algorithm \ref{alg:update_hop}, we define a threshold \(d_{\text{judge}} \) (line 4). When the distance between the empty bucket and the target position is below this threshold, we employ the single-step hopping method (lines 5-10), where each item is checked to see if it can be relocated. However, when the empty bucket is too far from the target position, sequentially checking each item would incur high computational overhead. 
In this case, we adopt a second method, the chain hopping approach (lines 11-19). This method checks each data item to determine its oldest version location by the \texttt{FrontInfo} (lines 12-14), and if the position is after the target location, we can move the entire data item backward (lines 15-17). This action does not violate any hop distance constraints, so no further validation is necessary. 
The chain hopping operation involves moving each data item to the position of its next item, with the last item being moved to the empty bucket (lines 24-31). 
This approach quickly brings the data item closer to the target position, thus reducing the number of comparisons.
The hop-bounded layout guarantees that the search scope for a specific version is restricted within a localized neighborhood in the hash bucket, preventing full-chain traversal even when version chains become long. This design preserves logarithmic retrieval time with respect to the number of versions per edge.

\noindent\textbf{Discussion.}
Although our design is inspired by the Hopscotch hashing algorithm, it differs fundamentally in its objective.
The original algorithm focuses on resolving hash collisions by using a bitmap to keep keys within a fixed neighborhood of their home bucket.
In contrast, our approach is designed to support ordered storage and efficient retrieval of multiple temporal versions of a single data item.
To this end, we replace the bitmap with item-centric \texttt{FrontInfo} and \texttt{BackInfo}, which explicitly capture inter-version relationships.
This design enables two novel mechanisms, \textit{single-step hopping} and \textit{chain hopping}, where the latter leverages these inter-version links and is unique to our structure.


\section{Temporal Query Engine}\label{sec:TEMPORAL_QUERY_ENGINE}
In this section, we present the query processes and time complexity analysis for Temporal Property Storage and Dynamic Topology Storage, and introduce the use of SIMD technology.

\subsection{Query Engine for Vertex Properties}\label{sec:1-1_QUERY_ENGINE}
To access the latest property version, the system directly reads from the relevant column using the vertex ID offset, avoiding version management overhead. For temporal queries at time t, it consults the vertex's \texttt{Temporal Table} and uses each \texttt{Version}'s \texttt{ModifyBitmap} and \texttt{Lifecycle} to locate the appropriate version entry via binary search. When multiple properties are queried simultaneously, parallel lookups are employed after the corresponding versions are identified to further accelerate retrieval. The actual data value's location is then determined by the \texttt{Offset}: if null, the current value is used; otherwise, the value is retrieved from the \texttt{Temporal Buffer}. Further, to optimize the requirement of finding all objects at the same time point, the system finds the first object as described above. Then, for all subsequent objects, it uses the \texttt{NextOffset} pointer from the previously queried \texttt{Version} to quickly locate their relevant \texttt{Version}, reducing comparison overhead and avoiding repeated full searches.
\begin{theorem}\label{NextOffset_proof}
Let the \texttt{NextOffset} of \(\texttt{Version}_i\) point to the next row's $\texttt{Version}_j$, with the lifecycles of $\texttt{Version}_i$ and $\texttt{Version}_j$ given by $[ti_{start}, ti_{end})$ and $[tj_{start}, tj_{end})$, respectively. 
Then, the lifecycles satisfy: 
\begin{equation}
[ti_{start}, ti_{end}) \cap [tj_{start}, tj_{end}) \neq \varnothing \land ti_{start} > tj_{start}
\end{equation}
\end{theorem}
Please see the proofs in our github~\cite{Sakuraaa0_CtGraph_2025}. This theorem allows us to start the version search for the next object from a temporally relevant position, effectively narrowing down the search space.

\noindent\textbf{I/O Analysis.} Here, we denote the number of vertices as \(N\), with an average version count per vertex represented by \(N_v\).

Read the latest version: Accessing the most current data is highly efficient in our system. It involves a direct lookup in the primary property columns, resulting in approximately one random access to fetch the data for the given object's property.

Read a specific historical version: To retrieve a historical version, \dbname first performs a random access to locate the entries in the \texttt{Temporal Table}, then uses binary search to find the correct \texttt{Version}. It finally retrieves the actual data from the \texttt{Temporal Buffer} using the \texttt{Offset}. This approach achieves a time complexity of $O(\log{N_v})$, compared to $O(N_v)$ for version chain methods. Moreover, since \texttt{Version} uses a contiguous array, \dbname only incurs random I/O at the final step, whereas version chain approaches require random I/O at each search step.

Read specific historical versions for all objects: 
When querying a full graph snapshot at time $t$, locating the first vertex's version is similar to a single-point query. 
For subsequent vertices, \dbname uses the \texttt{NextOffset} pointer to jump directly to the relevant region for each vertex, avoiding a full search. 
Finding the exact version within this region typically requires a small number $k$ of probes, leveraging temporal locality. 
Thus, after the initial search, the complexity for the remaining $(N-1)$ vertices is approximately $O((N-1)k)$, where $k$ is a small constant, resulting in an overall complexity of $O(N)$. 
In contrast, systems without this optimization may require $O(N \cdot N_v)$ time for $N$ independent full searches. Therefore, \texttt{NextOffset} substantially reduces the complexity of snapshot queries.

\subsection{Query Engine for Dynamic Topology}\label{sec:n-n_QUERY_ENGINE}

For cold data, since each vertex data is stored in a fixed, compact array, traversing the array can quickly locate the required version. 
For hot data in the \texttt{HopscotchHash Table}, the query process is as follows: to access the latest version of an edge, the system uses \texttt{LatestValuePtr} to directly locate the storage position by destination vertex ID, achieving $O(1)$ time complexity. For historical data at a specific time $t$, the system combines \texttt{HashFunction} and \texttt{LatestValuePtr} to determine the bounds of available versions, with storage distances capped by \texttt{MaxHopDistance}. Leveraging SIMD instructions, up to 16 slots can be compared simultaneously; when $\texttt{MaxHopDistance} \leq 16$, these versioned edges can be regarded as tightly stored, as validated in experiment~\ref{exper:Effectiveness of HopscotchTable}. How we leverage SIMD operations to accelerate query processing will be detailed in §~\ref{sec:simd}.
Overall, this enables fast, binary-search-style queries in a compact region, reducing time complexity from linear to logarithmic compared to traditional chained traversals. Additionally, for querying all objects at the same time point, \texttt{NextOffsetInfo} can be used to accelerate subsequent lookups, following a process similar to that described in §~\ref{sec:1-1_QUERY_ENGINE}.

\begin{table}
  \caption{Time Complexity of Different Operations in \dbname.}
  \label{tab:comparison}
  \resizebox{\linewidth}{!}{
\begin{tabular}{c|c|c|c}
     \hline
    \multirow{2}{*}{\textbf{Operation}} & \multicolumn{2}{c|}{\textbf{HopscotchHash Table}} & \multicolumn{1}{c}{\textbf{Temporal Table}}\\
    \cline{2-4}
     & \textbf{Situation} & \textbf{Time Complexity} & \textbf{Time Complexity}  \\
    \hline
    \multirow{2}{*}{Read the latest version} & $N_v \cdot N_e \leq S$ & $O(S)$ & \multirow{2}{*}{$O(1)$} \\
        & $N_v \cdot N_e > S$ & $O(1)$  &    \\
        \cline{1-4}
    Read a specific historical
      & $N_v \cdot N_e \leq S$ & $O(S)$  & \multirow{2}{*}{$O(\log(N_v))$} \\
     Version  & $N_v \cdot N_e > S$ & $O(\log(N_v))$  &   \\
        \cline{1-4}
   Read specific historical 
      & $N_v \cdot N_e \leq S$ & $O(S)$  & \multirow{2}{*}{$O(N)$}  \\
     versions for all objects & $N_v \cdot N_e > S$ & \(O(N \cdot N_e)\)  &   \\
        \cline{1-4}    
    \multirow{3}{*}{Insert, Update or Delete} 
      & $N_v \cdot N_e \leq S$ & $O(1)$  & \multirow{3}{*}{$O(1)$} \\
      & \multirow{1}{*}{$N_v \cdot N_e > S$} & 
      \begin{tabular}[c]{@{}c@{}}
        $\left\{\begin{array}{c} O(1)\ ...\ average \\ \Omega(N_v\cdot N_e)\ ...\ worst \end{array}\right.$
      \end{tabular}
        & \\
    \hline
  \end{tabular}
  }
\end{table}






\noindent\textbf{I/O Analysis.} Here, we denote the number of vertices as \(N\).
Each vertex in the SparseArray has a size of \( S \).
Each hash bucket can contain a maximum of $M$ items, and we represent the vertex degree as \( N_e \), \texttt{MaxHopDistance} as $H$, and the average number of versions per edge as \( N_v \).

Read the latest version: if \( N_v \cdot N_e \leq S \), the SparseArray contains fewer data items. A bounded scan of this small array, with complexity $O(S)$, is extremely efficient due to its excellent data locality.  When \( N_v \cdot N_e > S \), the data is stored in HopscotchHash Table, and the position of the most recent data item can be directly found using the \texttt{LatestValuePtr}, so the time complexity is \( O(1) \).

Read a specific historical version: SparseArray requires \(O(S)\), while HopscotchHash Table uses binary search with \(O(\log{N_v})\) due to sorted data and pointers.

Read historical versions for all objects: After identifying the first edge, subsequent versions can be found using \texttt{NextOffsetInfo}, resulting in \(O(N \cdot N_e)\) complexity.

Insert, Update or Delete: In SparseArray, insertion is \(O(1)\) due to direct access. In HopscotchHash Table, insertion is \(O(1)\) on average, but may require \(O(N_v \cdot N_e)\) in the worst case if resizing occurs. However, the hop failure probability is low, approximately \( \frac{1}{M \cdot H!} \), reducing this occurrence significantly.
\begin{theorem}\label{hop_proof}
The probability of hop failure in HopscotchHash Table is ~\( \frac{1}{M\cdot H!} \)
\end{theorem}
Please see the proofs in our github~\cite{Sakuraaa0_CtGraph_2025}.
Finally, Table~\ref{tab:comparison} summarizes the I/O characteristics of \dbname, focusing on its core components: the versioned ColumnStore with Temporal Table and the Hopscotch-optimized Hash Table.

\subsection{SIMD Acceleration}\label{sec:simd}

\begin{algorithm}[t]
\small
\caption{Get a Specific Temporal Edge}\label{alg:simd_functions}

\textbf{Input:} Search range start index $idx_{\text{start}}$, End index $idx_{\text{end}}$, Target edge $\mathcal{E}$, ControlInfo $controls$, HashBucket $\mathcal{B}$ \\
\textbf{Output:} The target edge's index\\

\LinesNumbered
\SetKwFunction{FGetVersionValue}{GetTemporalEdge}
\SetKwFunction{FMatch}{Match}
\SetKwFunction{FEdgeMatch}{EdgeMatches}
\SetKwProg{Fn}{Function}{:}{\KwRet}

\Fn{\FGetVersionValue{$idx_{\text{start}},\; idx_{\text{end}},\; \mathcal{E}$}}{
    \While{$idx_{\text{start}} \le idx_{\text{end}}$}{
        $idx_{\text{mid}} \leftarrow \lfloor (idx_{\text{start}} + idx_{\text{end}})/2 \rfloor$ \\
        $res_{\text{match}} \leftarrow$ \FMatch{$idx_{\text{mid}}, \mathcal{E}.control$} \\
        $upper\_bound\_updated \leftarrow \text{false}$ \\
        \While{$res_{\text{match}} \neq 0$}{
             $idx = \_\_builtin\_ctzll(res_{\text{match}}) + idx_{\text{mid}}$ \\
             $res_{\text{match}} \mathrel{\&}= res_{\text{match}} - 1$ \\
            \If{$\mathcal{B}[idx].dst = \mathcal{E}.dst$}{
                \uIf{$\mathcal{B}[idx].\tau \cap \mathcal{E}.\tau \neq \varnothing$}{
                    \Return $idx$ \\
                  }
                \ElseIf{$\mathcal{B}[idx].\tau.t_{\text{start}} > \mathcal{E}.\tau.t_\text{end}$}{
                    $idx_{\text{end}} \leftarrow idx - 1$\\
                    $upper\_bound\_updated \leftarrow \text{true}$ \\
                    \textbf{break}
                  }
            }
        }
       \If{not $upper\_bound\_updated$}{
        $idx_{\text{start}} \leftarrow idx_{\text{mid}} + 16$\\
    }
    }
    \Return IDX\_NOT\_FOUND
}

\Fn{\FMatch{$i,\; c$}}{
    $v \leftarrow \_mm\_loadu\_si128(\text{controls} + i)$ \\
    $w \leftarrow \_mm\_set1\_epi8(c)$ \\
    $x \leftarrow \_mm\_cmpeq\_epi8(v,\; w)$ \\
    $mask_{\text{match}} \leftarrow \_mm\_movemask\_epi8(x)$ \\
    \Return $mask_{\text{match}}$\\
}
\end{algorithm}

To further improve query performance, \dbname leverages Single Instruction, Multiple Data (SIMD) techniques. Next, we elaborate in detail how we utilize this technique to get an edge in our HopscotchHash Table.
As discussed in §~\ref{Hopscotch Hash}, the control array \texttt{ControlInfo} records the upper seven bits of the hash value along with a control bit, forming an 8-bit control field. Specifically, values from 0x00 to 0x7F indicate a valid entry and encode the upper seven hash bits; 0x80 denotes an empty slot; and 0xFF marks a deleted entry.
This design allows for an initial assessment of whether a candidate slot may contain the target data item by simply comparing the 8-bit control field during query operations. Since each control field occupies only 1 byte, its small size allows it to be efficiently loaded into the cache, which reduces memory access latency.

We implement a SIMD-optimized binary search process within the \texttt{GetTemporalEdge} function, as shown in Algorithm~\ref{alg:simd_functions}.
Instead of directly comparing full data items, it first calls the \texttt{Match} function (line 6) to rapidly filter a batch of candidates by matching their Control Fields.
The \texttt{Match} function is where SIMD is applied (lines 21-26). It first loads 16 Control Fields into a SIMD register (line 22). 
Then, another instruction \texttt{\_mm\_cmpeq\_epi8} performs a parallel comparison of these 16 bytes against the target hash's upper bits (line 24). 
The \texttt{\_mm\_movemask\_epi8} instruction efficiently converts the 16-byte comparison result into a 16-bit bitmap, identifying all matching candidates at once (line 25). 
After that, we compare the data items matched by the \texttt{Match} function (lines 8-17).

The application of SIMD technology in Temporal Property Storage lies in \texttt{Version} lookup, with the specific operational procedure being largely consistent with the above description. 
This approach allows \dbname to perform an initial screening on multiple data items using just a few CPU instructions. It reduces the latency associated with traditional item-by-item comparisons and accelerates the location and retrieval of historical versions.

\section{Implementation}\label{Imple}

\dbname is implemented in C++ with $\sim$7,000 lines of code. 
\dbname realizes the hybrid architecture described above: it combines versioned columnar storage with an optimized Hopscotch hash table to enable efficient storage and querying of temporal graphs. \dbname supports both property and non-property graph input formats. When the input is a non-property graph, \dbname only utilizes Dynamic Topology Storage for the entire graph. \dbname implements all temporal graph data management operations defined in §~\ref{sec:temporal_graph_model}, providing users with rich interfaces to perform complex queries and updates directly on temporal graphs.

\dbname supports Snapshot Isolation (SI). To ensure data consistency in concurrent environments, \dbname uses fine-grained locking at the \texttt{Header} of each object’s
\texttt{Temporal Table} and \texttt{HopscotchHash Table}, enabling thread-safe parallel updates and queries. This locking granularity avoids the performance bottlenecks caused by global locks, improves data access efficiency under multi-threaded workloads, and guarantees both atomicity and consistency of operations. 


\noindent\textbf{Fault Tolerance and Persistence.}
To ensure data durability, TVA implements a Write-Ahead Logging (WAL) mechanism. Each WAL entry is structured as a compact binary sequence starting with a header that contains a 4-byte magic number, version identifier, operation type and flags. This header is followed by an 8-byte timestamp, payload length, the variable-length payload, and finally a CRC32 checksum to guarantee data integrity during recovery. To address diverse storage constraints, TVA introduces a flexible persistence architecture managed via a \texttt{PersistManager} that supports two distinct modes. The \textit{Pointer Mode}, designed for high-frequency access, retains TVA's core in-memory structures (i.e., \texttt{ColumnTable}) to preserve traversal speed, but replaces bulky property values with lightweight 64-bit \texttt{DiskOffset} pointers referencing an append-only disk file. For scenarios requiring full persistence, the \textit{Full Mode} leverages RocksDB as the backend. We design a specialized time-encoded key schema, specifically encoding keys as \texttt{user:\{uid\}:\allowbreak ts:\{timestamp\}} for vertices and \texttt{friend:\{src\}:\allowbreak \{dst\}:\allowbreak ts:\{timestamp\}} for edges, to exploit the lexicographical ordering of the underlying LSM-tree. This ensures that historical versions are physically adjacent on disk, enabling efficient point-in-time queries via iterator seeking and temporal range scans without loading the entire history into memory.

\section{Evaluation}\label{sec:EVALUATION}

In this section,
we evaluate \dbname through comparisons with state-of-the-art graph systems.
We conduct system-level and micro benchmarks to measure the query performance and storage consumption.
\subsection{Experiment Setup}\label{sec:setup}

We conduct our experiments on a machine equipped with an Intel(R) Xeon(R) Gold 5220 CPU @ 2.20GHz and 128 GB of memory, running Ubuntu 20.04. 
All code is compiled using GCC 11.3.0 with the \texttt{O3} optimization flag. 
All systems utilize 5 threads by default.

\noindent\textbf{Competitors.}
We compare \dbname with two categories of graph storage systems.
1) For temporal data operations, we compare with the most advanced graph data structures that support temporal functionality: Clock-G~\cite{DBLP:conf/icde/MassriMPM22}, T-GQL~\cite{DBLP:journals/vldb/DebrouvierPPSV21}, and AeonG~\cite{DBLP:journals/pvldb/HouZWLJWD24}. We also included PostgreSQL~\cite{postgresql_temporal} (representing temporal-extended RDBMS) and RocksDB~\cite{rocksdb_github} (representing Key-Value stores) in our experiments.
Note that both Clock-G and T-GQL store all vertices and edges in memory. To ensure a fair comparison, we cache all data in memory for all the evaluations.
2) For recent data operations, we select state-of-the-art dynamic graph storage systems as baselines for comparison: GraphOne~\cite{DBLP:conf/fast/KumarH19}, Stinger~\cite{DBLP:conf/hpec/EdigerMRB12}, and Sortledton~\cite{DBLP:journals/pvldb/FuchsGM22}. 
Since these structures do not support temporal operation.
To ensure a fair comparison for MVCC graph stores, we standardize the experimental setup by maintaining complete version histories across all compared systems to rigorously evaluate their temporal query capabilities without the interference of garbage collection.

\noindent\textbf{Graph Datasets.}
We employ a variety of graph datasets for evaluation, including: IMDB~\cite{nr}, an actor collaboration network from IMDB; DBLP~\cite{DBLP:conf/icdm/YangL12}, a collaboration network based on DBLP; YouTube \cite{DBLP:conf/icdm/YangL12}, a social network from YouTube; Epinions~\cite{DBLP:conf/trustcom/SaccoBD13}, the Epinions ``who-trust-whom'' social network; Pokec~\cite{DBLP:journals/corr/abs-2208-14807}, Slovakia’s most popular social network; and LDBC SNB~\cite{DBLP:conf/sigmod/ErlingALCGPPB15}, the LDBC Social Network Benchmark designed to simulate real-world social networks.

\noindent\textbf{Graph Benchmark.}
For temporal queries, we utilize three temporal workloads: T-mgBench, T-LDBC, and T-gMark, following~\cite{DBLP:journals/pvldb/HouZWLJWD24}.
T-mgBench is based on the Pokec dataset and follows the Memgraph mgBench workload.
Specifically, the ``\texttt{FOR TT AS OF} $t$'' clause is appended to queries Q1 and Q3 to generate ``time point'' queries, while the ``\texttt{FOR TT FROM} $t_1$ \texttt{TO} $t_2$'' clause is included in Q2 and Q4 to produce temporal range queries. 
T-LDBC is adapted from the LDBC workload by integrating ``\texttt{FOR TT AS OF} $t$'' into the IS queries (IS1–IS7).
T-gMark is based on gMark~\cite{DBLP:journals/tkde/BaganBCFLA17} and is designed to stress-test property management. It features 24 vertex properties and 82 edge properties, covering diverse query shapes and selectivities.





For recent data operations, we consider four representative graph analysis algorithms, including Breadth-First Search (BFS), Single-Source Shortest Path (SSSP), PageRank (PR), and Weakly Connected Components (WCC). 


\subsection{Experiment on Temporal Graph}\label{sec:tg}
\subsubsection{Experiments on temporal graph storage consumption.}

\begin{figure}
  \centering
  \includegraphics[width=\linewidth]{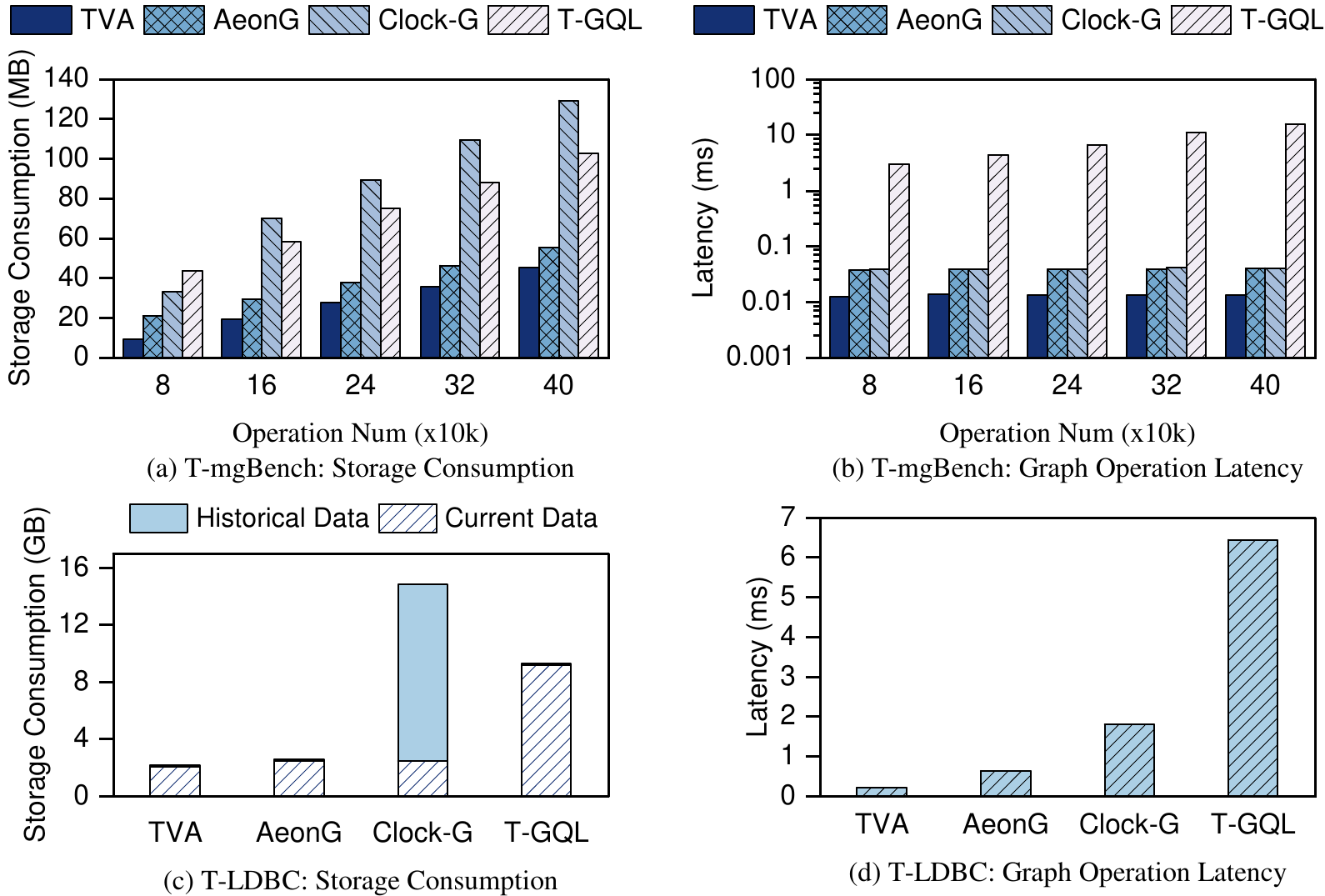}
  \caption{Temporal Graph Storage Consumption and Graph Operation Latency.}
  \Description{Comparisons on Storage Consumption and Graph Operation Latency.}
  \label{fig:ex_T_MgBenchOperation}
\end{figure}

\begin{figure*}
  \centering
  \includegraphics[width=\linewidth]{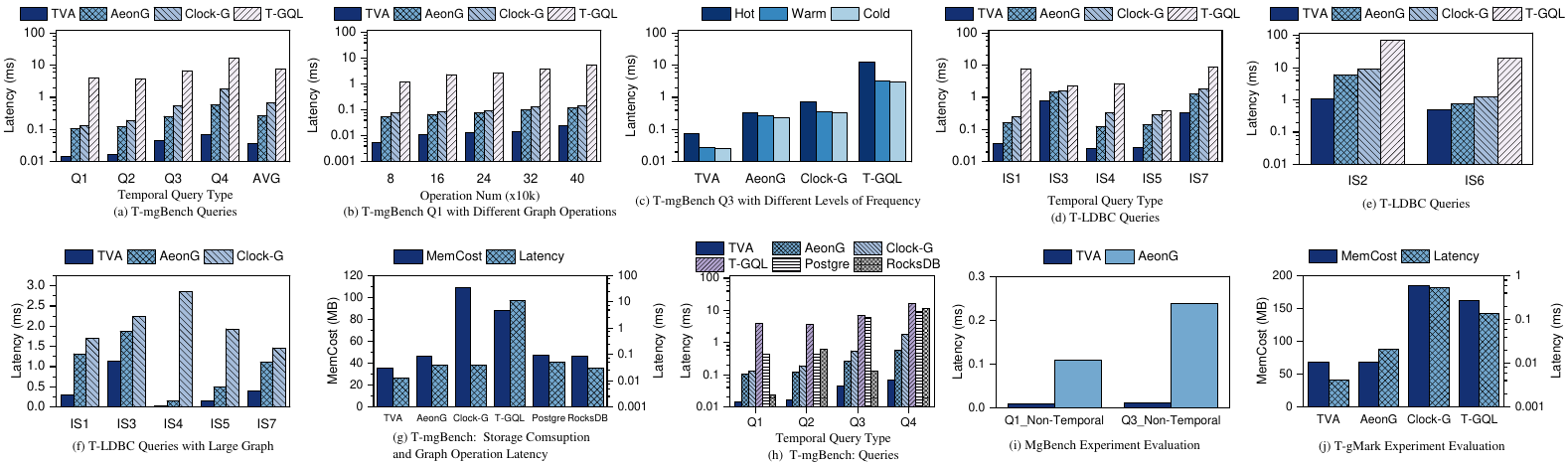}
  \caption{Comparisons on Temporal Query Latency.}
  \Description{Comparisons on Temporal Query Latency.}
  \label{fig:ex_History_Graph_Query}
\end{figure*}

We first use T-mgBench to analyze storage consumption under varying numbers of graph operations, as shown in Figure~\ref{fig:ex_T_MgBenchOperation}(a). 
The results show that \dbname demonstrates the lowest storage overhead across different operation volumes. 
\dbname's storage efficiency surpasses AeonG by up to $2.2\times$, Clock-G by up to $3.6\times$, and T-GQL by up to $4.7\times$. 
These storage savings primarily result from \dbname’s compact storage structure. The system efficiently organizes historical data in the Temporal Buffer and employs a low-overhead structure to manage version metadata, thereby minimizing the overhead associated with maintaining historical data. In contrast, AeonG periodically stores full data within the version chain; T-GQL maintains complete information for all data; and Clock-G periodically creates full historical snapshots of the graph. All these approaches lead to higher storage overhead compared to \dbname.

The results on the T-LDBC dataset, shown in Figure~\ref{fig:ex_T_MgBenchOperation}(c), further demonstrate this advantage and provide a clearer breakdown of storage composition. 
\dbname achieves up to $1.2\times$, $6.7\times$, and $4.3\times$ lower storage consumption than AeonG, Clock-G, and T-GQL, respectively. 
The total storage required by \dbname remains minimal, with the ``Historical Data'' component being notably small in comparison to the ``Current Data''.
In contrast, for Clock-G, the Historical Data constitutes a large proportion of the total storage. 
These results indicate that Clock-G's approach of periodically creating full historical snapshots substantially increases storage overhead.

\subsubsection{Experiments on temporal graph operation latency.}
To evaluate the performance of temporal graph operations, we measured the average latency on the T-mgBench and T-LDBC, with the results presented in Figure \ref{fig:ex_T_MgBenchOperation}(b) and Figure \ref{fig:ex_T_MgBenchOperation}(d).

Figure \ref{fig:ex_T_MgBenchOperation}(b) shows that on T-mgBench, \dbname consistently maintains the lowest and most stable average graph operation latency as the number of operations increases from 80k to 400k. In contrast, other systems exhibit increased high latency. At 400k operations, \dbname's average latency is $3\times$, $3.1\times$, and $1197.6\times$ faster than AeonG, Clock-G, and T-GQL. This advantage is primarily due to \dbname's efficient historical data management. Specifically, \dbname avoids the overhead caused by AeonG’s approach of archiving the latest data to a historical storage area and recording changes via a version chain, which introduces additional computational cost. Meanwhile, T-GQL suffers from inefficient traversal because it does not distinguish between current and historical data.

Subsequent experiments on the larger and more complex T-LDBC dataset, shown in Figure \ref{fig:ex_T_MgBenchOperation}(d), further corroborate \dbname's low-latency advantage.
\dbname records the lowest average operation latency, with its performance being $2.9\times$ faster than AeonG, $8.1\times$ faster than Clock-G, and $28.9\times$ faster than T-GQL. 
Notably, on the large-scale T-LDBC graph, Clock-G's periodic creation of large historical snapshots consumes significant CPU and I/O resources. 
This resource contention adversely impacts its real-time graph operation performance, explaining its relatively higher latency.

\subsubsection{Experiments on temporal graph analysis performance.}

We first evaluate temporal query performance using T-mgBench (Q1–Q4). As shown in Figure~\ref{fig:ex_History_Graph_Query}(a)-(c), \dbname consistently outperforms AeonG, Clock-G, and T-GQL across all queries, achieving average latency reductions of up to $206.9\times$. This advantage stems primarily from our \texttt{Temporal Table} enabling efficient direct access to historical data, and our Hopscotch-based algorithm ensuring contiguous storage to minimize unnecessary scans. In contrast, AeonG suffers from traversing long version chains, Clock-G incurs overhead from aggregating multiple snapshots, and T-GQL performs expensive full-graph traversals. Furthermore, \dbname maintains this superior performance under varying workloads, outperforming competitors even as the volume of graph operations scales up.

To validate the effectiveness of \dbname in larger-scale and more complex scenarios, we conducted evaluations on T-LDBC. As shown in Figures \ref{fig:ex_History_Graph_Query}(d) and (e), \dbname continues to demonstrate superior performance across various query types, achieving latency reductions of up to $4.4\times$, $6.6\times$, and $207.1\times$ compared to AeonG, Clock-G, and T-GQL, respectively. 
To further stress-test the systems, we evaluated an even larger graph containing 9.28 million vertices and 52.7 million edges. During this experiment, T-GQL failed to complete the test due to an Out-Of-Memory error. The results for the remaining systems are shown in Figure \ref{fig:ex_History_Graph_Query}(f), where \dbname achieves performance improvements of up to $4.7\times$ and $83.9\times$ over AeonG and Clock-G.

To justify the necessity of a specialized temporal graph store over general-purpose systems, we compared TVA against PostgreSQL (representing temporal-extended RDBMS) and RocksDB (representing Key-Value stores). The results in Figure~\ref{fig:ex_History_Graph_Query}(g) and (h) reveal fundamental architectural bottlenecks in these baselines. PostgreSQL relies on expensive join operations to traverse edges, resulting in a combinatorial explosion of intermediate results and a sharp decline in performance for multi-hop temporal queries. By utilizing a version-aware storage layout that replaces costly joins with direct pointer dereferencing, TVA achieves a $126.4\times$ performance improvement over PostgreSQL. Similarly, although RocksDB is efficient for simple point queries, it lacks structural awareness, leading to excessive random I/O and inefficient scans during the multi-hop traversals and temporal range scans required by time-series graph workloads.

To address concerns about system overhead in non-\allowbreak temporal scenarios, we compared the performance of TVA and baseline systems on mgBench with all temporal data excluded. The results in Figure~\ref{fig:ex_History_Graph_Query}(i) indicate that while all systems naturally experience a performance boost due to the elimination of version retrieval steps, TVA maintains its superior architectural efficiency, achieving a $21.7\times$ performance improvement over AeonG on static snapshots.

Furthermore, to evaluate property-intensive workloads on T-gMark, Figure~\ref{fig:ex_History_Graph_Query}(j) shows TVA's architecture decouples property values from version metadata, enabling parallel retrieval of multiple properties once the version metadata is located in the \texttt{Temporal Table}. This yields speedups of $5.1\times$, $132.9\times$, and $33.4\times$ over AeonG, Clock-G, and T-GQL, respectively.

\subsection{Experiment on Current Graph}\label{sec:cg}

\begin{figure}
  \centering
  \includegraphics[width=\linewidth]{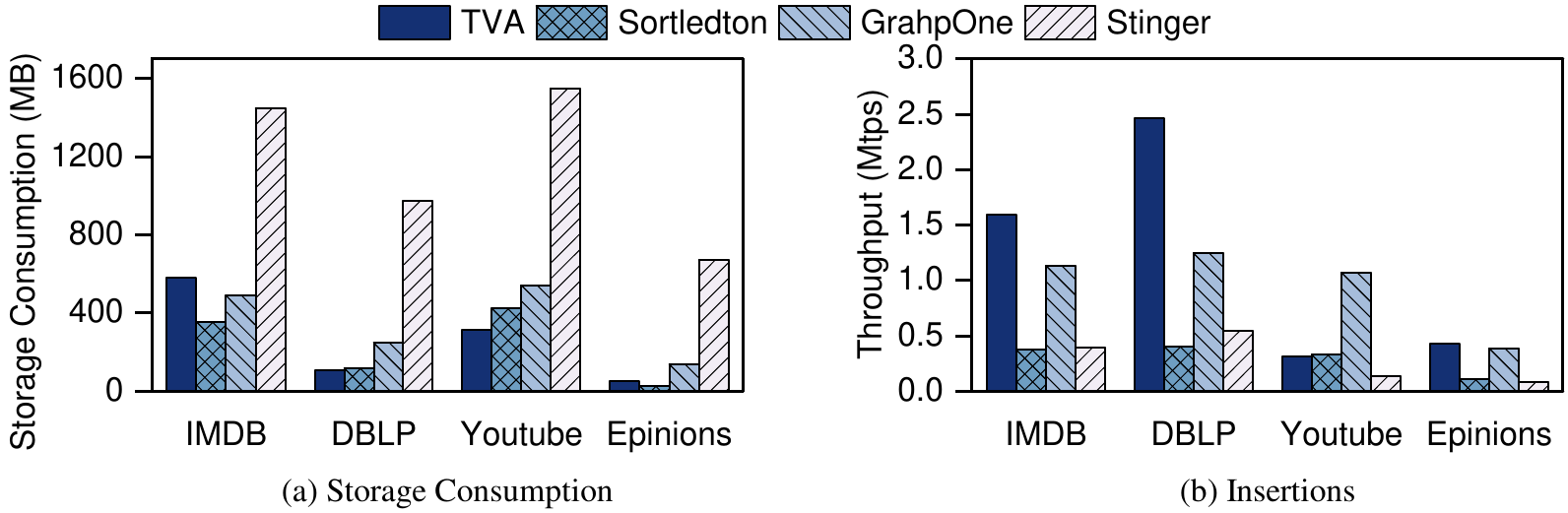}
  \caption{Current Graph Storage Consumption and Graph Operation Throughput.}
  \Description{the average throughput and  memory consumption measured in the insertions among various graphs.}
  \label{fig:ex_randomInsert}

\end{figure}

\subsubsection{Experiments on current graph storage consumption.}

We first evaluate the current data storage overhead of \dbname. 
The results are shown in Figure \ref{fig:ex_randomInsert}(a).
The results demonstrate that \dbname exhibits superior storage efficiency on the majority of datasets. Compared to the baseline systems, \dbname’s space utilization is up to $1.4\times$, $1.8\times$, and $5.1\times$ higher than Sortledton, GraphOne, and Stinger, respectively. 
An exception occurs on the IMDB and Epinions datasets, where \dbname’s storage overhead is slightly higher than that of Sortledton. 
This is attributed to \dbname’s storage management strategy, which pre-allocates fixed-size space for low-degree vertices. 
When the graph contains a large number of vertices with very low degrees, this strategy can lead to some wasted space. 
We consider this overhead acceptable because the reserved space is a key design choice that enables high insertion throughput for low-degree vertices.

\subsubsection{Experiments on current graph operation latency.}


We then evaluate the insertion performance of different graph systems. Figure \ref{fig:ex_randomInsert}(b) shows the insertion throughput across multiple datasets. 
On the YouTube dataset, GraphOne marginally outperforms \dbname, mainly because most edges connect to a few high-degree vertices, causing more hash collisions during position lookups. Notably, GraphOne skips edge existence checks in this experiment; enabling such checks would reduce its throughput to about 5 edges per second~\cite{DBLP:journals/pvldb/LeoB21}, due to the overhead of snapshot creation.
On all other datasets, \dbname consistently achieves up to $6.1\times$, $2\times$, and $4.5\times$ higher insertion throughput than Sortledton, GraphOne, and Stinger. 
This can be attributed to \dbname's hybrid insertion approach: it pre-allocates space for low-degree vertices to avoid frequent memory operations and utilizes a HopscotchHash Table for fast indexing of high-degree vertices, supporting efficient edge insertions. 
These features highlight \dbname’s adaptability to diverse graph structures.

\subsubsection{Experiments on current graph query latency.}

We further evaluate the performance of typical graph analysis tasks by measuring the execution times of four graph analysis algorithms, and compare \dbname with its competitors. Figure \ref{fig:ex_currentAnalyze} presents the normalized execution times of different systems across various datasets, where the execution time of \dbname is set as the baseline and normalized to 1. 
\dbname outperforms other methods in most cases, achieving speedups of up to $6.5\times$, $18.4\times$, and $5.6\times$ compared to Sortledton, GraphOne, and Stinger. 
GraphOne suffers the largest performance decline during graph analysis due to its block-based adjacency list operations, which cause substantial random access.
Sortledton outperforms others in BFS because its skiplist structure makes it convenient to access all neighbors of a vertex, while \dbname incurs extra overhead from hash table lookups during edge traversal.
However, this approach is less effective for algorithms such as PR and SSSP that require frequent access to neighbors of different vertices.

\begin{figure}
  \centering
  \includegraphics[width=\linewidth]{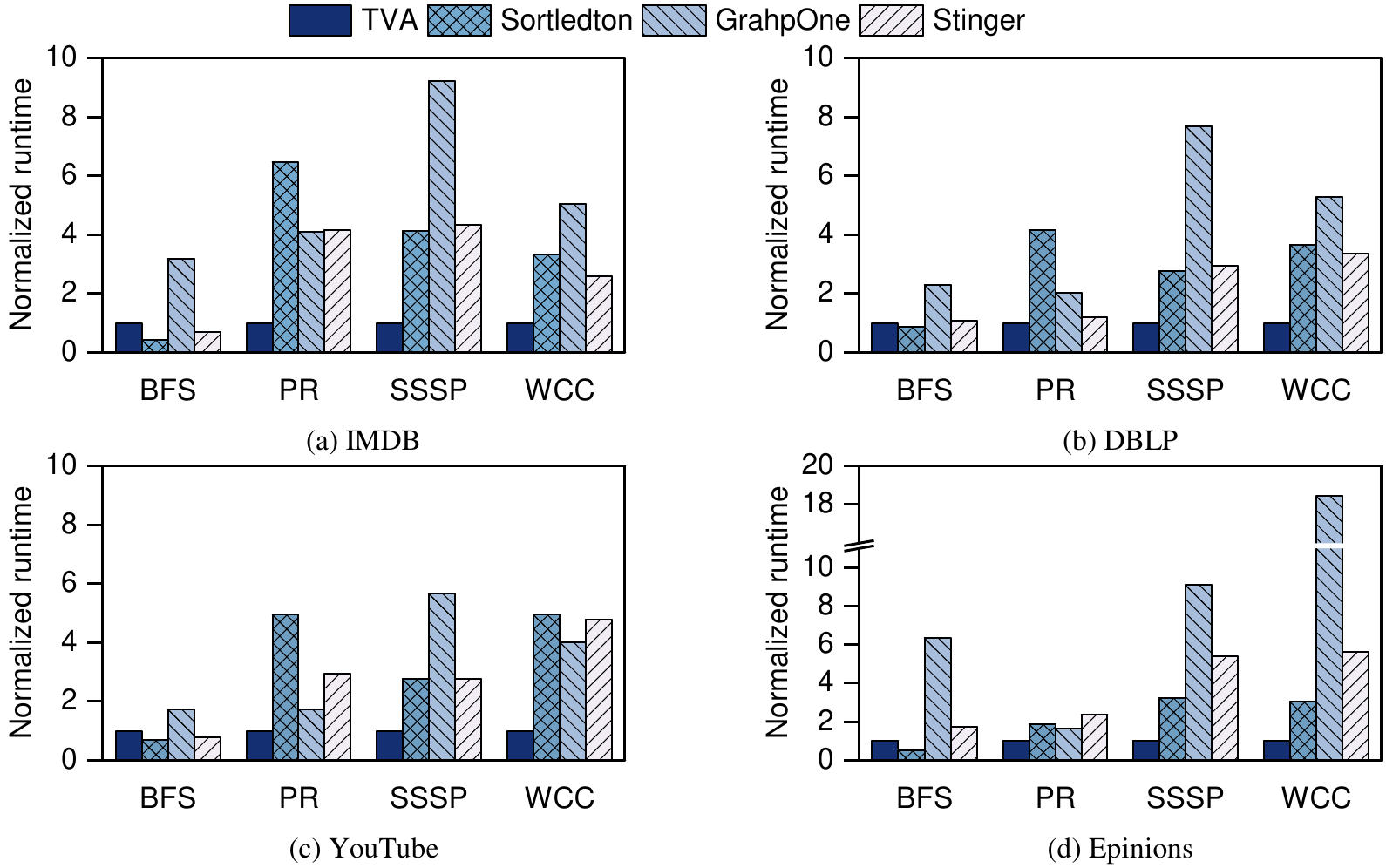}
  \caption{Current Graph Query Latency.}
  \Description{The execution time of graph algorithms on different datasets.}
  \label{fig:ex_currentAnalyze}
\end{figure}
\subsection{Persistence Analysis}\label{sec:persist}

\subsubsection{Effectiveness of TemporalChain in Long Version Chains.}
We evaluate TemporalChain for querying items at a specific timestamp. In Figure~\ref{fig:ex_consist}(a), TemporalChain achieves up to $6.2\times$ speedup over independent item lookups by leveraging inter-item links to directly navigate to subsequent relevant versions. 
By confining the search scope within a bounded neighborhood and utilizing TemporalChain to bypass repeated full-chain traversals, \dbname effectively circumvents the linear performance degradation ($O(N_v)$) typical of traditional linked-list version-chain approaches. 
Our design ensures that version retrieval time remains logarithmic ($O(\log N_v)$), making \dbname's computational advantage increasingly prominent under heavy, history-intensive workloads.

\subsubsection{Persistence Overhead.} 
To rigorously quantify the cost of durability and I/O overhead when datasets exceed memory capacity, we evaluate \dbname under two distinct memory-spillover persistence strategies managed by the PersistManager: TVA\_1 (Pointer Mode) and TVA\_2 (Full Mode).
Figures~\ref{fig:ex_consist}(b) and (c) detail the storage footprint and operation latency. The Pointer Mode retains lightweight topology metadata in RAM while offloading bulky historical property values to an append-only disk file via 64-bit offsets, thus reducing the memory footprint. 
The Full Mode fully persists both topology and properties by leveraging RocksDB as a backend, utilizing a specialized time-encoded key schema (e.g., \texttt{user:uid:ts:timestamp}) to exploit LSM-tree lexicographical ordering so that historical versions remain physically adjacent on disk. 
Furthermore, we explicitly measured the overhead of our Write-Ahead Logging (WAL) mechanism. By serializing both version metadata and actual data into a highly compact binary log format—featuring a lightweight header and CRC32 checksum for data integrity—our persistence mechanism introduces only marginal latency penalties.
The results confirm \dbname maintains robust, competitive throughput even under strict durability guarantees.

\begin{figure}
  \centering
  \includegraphics[width=\linewidth]{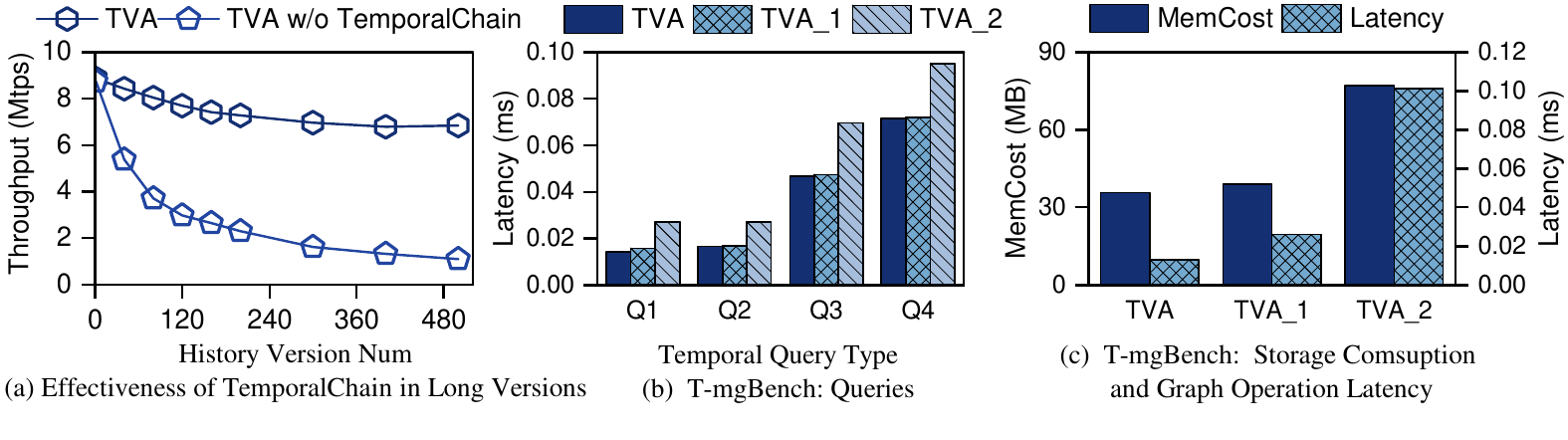}
  \caption{Persistence Analysis on \dbname.}
  \Description{Consistency Analysis on \dbname.}
  \label{fig:ex_consist}

\end{figure}

\subsection{Ablation Study on \dbname}

\subsubsection{Effectiveness of HopscotchTable.}\label{exper:Effectiveness of HopscotchTable}
We evaluate the HopscotchHash Table's hop distance and query efficiency. Figure~\ref{fig:ex_xiaorong}(a) shows that under varying Zipf skewness ($\alpha$), the hop distance for 99\% of data insertions peaks at 8 and strictly remains below 16.
This ensures versions are localized within a bounded, SIMD-friendly neighborhood.
Compared to a standard MVCC version-chain baseline~\cite{DBLP:conf/sigmod/KimKCYKJ21}, Figure~\ref{fig:ex_xiaorong}(b) demonstrates that HopscotchHash Table achieves up to $2.9\times$ speedup as the history length grows, reducing lookup complexity to $O(\log N_v)$ via binary search.

\subsubsection{Effectiveness of SIMD} 
As shown in Figure~\ref{fig:ex_xiaorong}(c), leveraging SIMD instructions provides up to $1.4\times$ query speedup by enabling parallel 16-byte candidate filtering. 
While the relative performance gap narrows with longer historical version chains and larger data volumes, SIMD remains a valuable optimization for rapid candidate screening when items have fewer historical versions. 

\subsubsection{Effectiveness of Migration Mechanism.}
We migrate a vertex to the HopscotchHashTable when either its degree exceeds $T_{deg}$ or its update frequency exceeds $T_{ver}$; Figure~\ref{fig:ex_xiaorong}(d) identifies the optimal boundary via normalized runtime across varying degrees. To avoid latency penalties, migration runs asynchronously: the triggering write flags the vertex and completes without blocking, while a background worker allocates a new bucket, re-organizes the edges, and atomically updates the pointer. 

\subsubsection{Scalability and Update Performance.} 
To explicitly demonstrate \dbname's concurrency and scalability under update-intensive conditions, we evaluated mixed read-write workloads. Figure~\ref{fig:ex_xiaorong}(e) illustrates that \dbname achieves near-linear throughput scaling as thread counts increase.
Furthermore, Figure~\ref{fig:ex_xiaorong}(f) utilizes varying write-read ratios to simulate realistic update-intensive workloads in concurrent operations.
The results verify that our fine-grained Snapshot Isolation (SI) implementation, which locks only at the individual object's Temporal Table Header rather than globally, effectively minimizes lock contention. 
This allows \dbname to maintain high query throughput without blocking readers, even under severe update pressure.

\begin{figure}
  \centering
  \includegraphics[width=\linewidth]{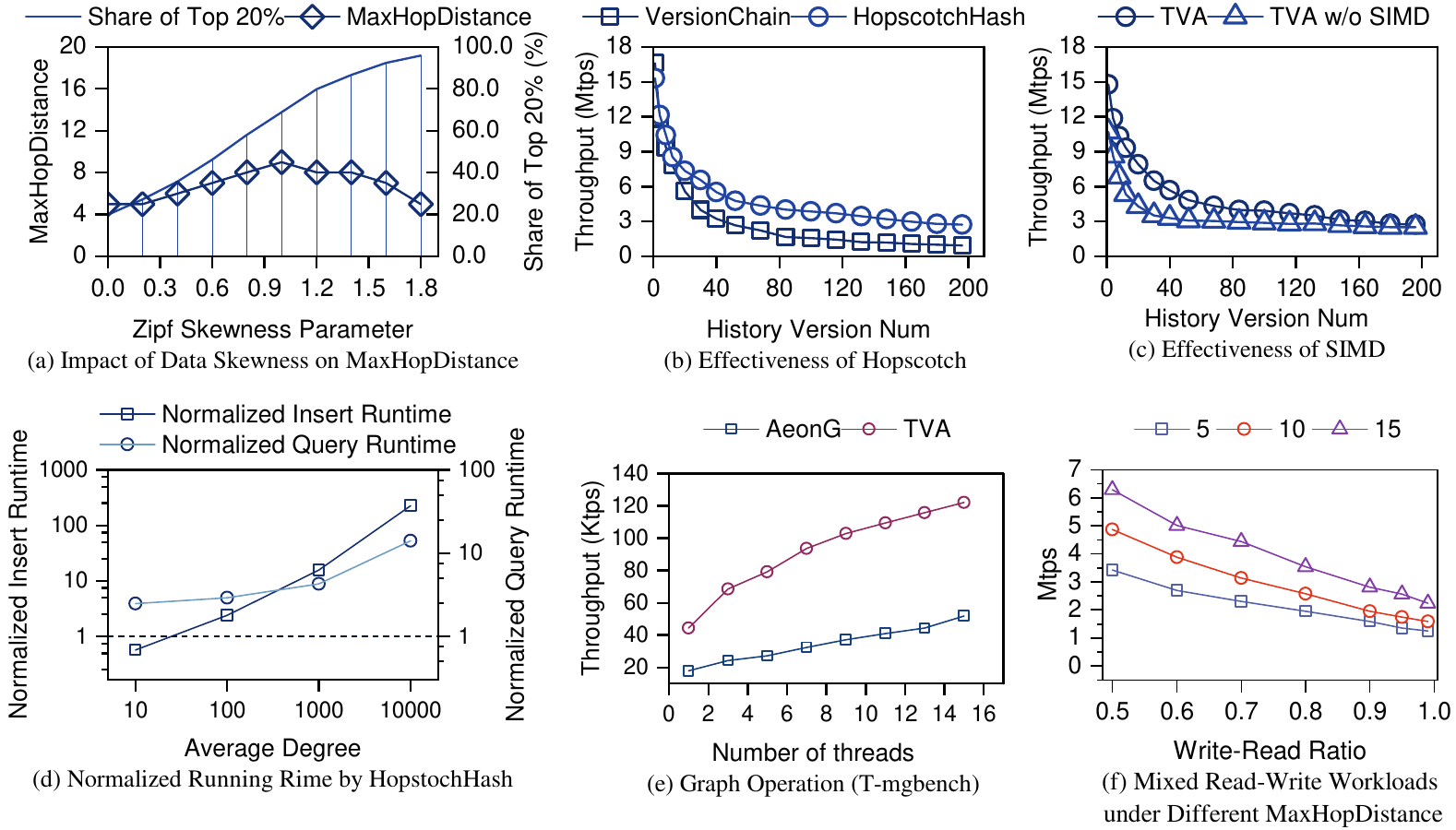}
  \caption{Performance Analysis on \dbname.}
  \Description{Performance Analysis on \dbname.}
  \label{fig:ex_xiaorong}
\end{figure}

\section{Related Works}\label{sec:RELATED WORKS}

\noindent\textbf{Temporal Graph Management.}
Recently, many works focus on managing and querying temporal graph data. AeonG~\cite{DBLP:journals/pvldb/HouZWLJWD24} extends graph databases with temporal support, but it cannot completely address the disadvantages posed by version chain scanning. Snapshot-based systems such as T-GQL~\cite{DBLP:journals/vldb/DebrouvierPPSV21} provide clear historical models but suffer from high storage costs and inefficient query performance due to full snapshot reconstruction. Kineograph~\cite{cheng2012kineograph} also explores fast-changing graphs, but with a focus on distributed environments and only capturing recent changes.

\noindent\textbf{Multi-Version Database Systems.} Classic multi-version database systems (MVCC) were primarily designed for high-concurrency transactional workloads, where historical versions are side effects of isolation rather than a persistent analytical asset~\cite{DBLP:conf/sigmod/KimKCYKJ21}. Consequently, MVCC systems typically assume version chains are short and rely on aggressive garbage collection or vacuuming to remove obsolete versions and control storage growth and lookup overhead~\cite{DBLP:conf/osdi/Zhang0Y24,DBLP:journals/pvldb/WuALXP17}. 
In contrast, temporal graph databases must preserve long histories for time snapshot and temporal-range analytics, and their workloads further stress version management, making strong temporal locality across scans critical.

\textbf{General-purpose Storage Systems.} 
Temporal graphs can be mapped to relational tables such as vertex and edge tables and queried with relational operators, and modern RDBMSs also provide temporal features~\cite{rocksdb_github,postgresql_temporal}. However, even with temporal extensions, relational execution for multi-hop traversals typically requires repeated joins, which can cause a combinatorial blow-up as hop count increases.
Likewise, KV stores excel at point lookups but lack structural awareness of graph topology. As a result, multi-hop traversals and temporal range scans suffer from excessive random accesses and poor locality.

\section{Conclusion}\label{sec:CONCLUSION}

This paper presented \dbname, an efficient temporal graph storage system.
\dbname decoupled version metadata from actual data, and based on this principle, we designed tailored data structures, i.e., the
temporal table and the enhanced hopscotch-based hash table, to enable fast version retrieval.
Further, we introduced a version-skipping approach and leveraged SIMD parallelism to support high-performance temporal query processing.
Experimental results demonstrated that \dbname achieves superior performance in terms of temporal query latency, update efficiency, and storage compactness compared to state-of-the-art graph storage systems.

\begin{acks}
This work is supported by the National Natural Science Foundation of China under Grant 62441230, 62461146205, 62072458 and 62472429.
\end{acks}

\balance
\bibliographystyle{ACM-Reference-Format}
\bibliography{ref}

\end{document}